\documentclass[sn-mathphys]{sn-jnl}% Math and Physical Sciences Reference Style
\jyear{2021}%

\theoremstyle{thmstyleone}%
\newtheorem{theorem}{Theorem}%  meant for continuous numbers
\theoremstyle{thmstyletwo}%
\newtheorem{example}{Example}%
\newtheorem{remark}{Remark}%

\theoremstyle{thmstylethree}%

\usepackage{color}
\usepackage{enumerate}
\usepackage{multirow}

\def\hatn{\hat{\mathbf{n}}}

\def\hatn{\hat{\mathbf{n}}}
\def\hatp{\hat{\mathbf{p}}}
\def\hatq{\hat{\mathbf{q}}}

\def\hatP{\hat{\mathbf{P}}}
\def\hatQ{\hat{\mathbf{Q}}}
\def\rmd{\mathrm{d}}
\def\hatT{\hat{\mathbf{T}}}
\def\hatN{\hat{\mathbf{N}}}
\def\hatB{\hat{\mathbf{B}}}

\def\bfr{\mathbf{r}}
\def\bfR{\mathbf{R}}

\begin{document}

\title{Compatible Director Fields in $\mathbb{R}^3$}


\author*[1]{\fnm{Luiz C. B.} \sur{da Silva}}\email{luiz.da-silva@weizmann.ac.il}

\author[1]{\fnm{Tal} \sur{Bar}}\email{tal.bar@weizmann.ac.il}

\author*[1]{\fnm{Efi} \sur{Efrati}}\email{efi.efrati@weizmann.ac.il}

\affil[1]{\orgdiv{Department of Physics of Complex Systems}, \orgname{Weizmann Institute of Science}, \orgaddress{\city{Rehovot} \postcode{7610001}, \country{Israel}}}

% \affil[2]{\orgdiv{Department}, \orgname{Organization}, \orgaddress{\street{Street}, \city{City}, \postcode{10587}, \state{State}, \country{Country}}}

% \affil[3]{\orgdiv{Department}, \orgname{Organization}, \orgaddress{\street{Street}, \city{City}, \postcode{610101}, \state{State}, \country{Country}}}


\abstract{The geometry and interactions between the constituents of a liquid crystal, which are responsible for inducing the partial order in the fluid, may locally favor an attempted phase that could not be realized in $\mathbb{R}^3$. While states that are incompatible with the geometry of $\mathbb{R}^3$ were identified more than 50 years ago, the collection of compatible states remained poorly understood and not well characterized. Recently, the compatibility conditions for three-dimensional director fields were derived using the method of moving frames. These compatibility conditions take the form of six differential relations in five scalar fields locally characterizing the director field. In this work, we rederive these equations using a more transparent approach employing vector calculus. We then use these equations to characterize a wide collection of compatible phases.\\

\textbf{Note.} The final publication is available in the ``Journal of Elasticity" via 

\url{https://doi.org/10.1007/s10659-023-09988-7}}

\keywords{Geometric frustration, Incompatibility, Liquid crystal, Frobenius}



\maketitle

\section{Introduction}
\label{sec:intro}

{Liquid crystalline textures are often described by extrinsic quantities. For example, a distorted nematic phase will be described using the polar and azimuthal angles describing the orientation of the unit director field relative to some fixed Cartesian frame. In contrast, the short-range interactions that give rise to the liquid crystalline phase and its resulting texture prescribe the relative positioning and orientation of the constituents {\em locally}. To properly describe these locally preferred relative positioning and orientation of the constituents, one must resort to an intrinsic description, i.e., a description by quantities available to an observer residing within the material manifold (and oblivious of the lab frame of reference). Examples of such local intrinsic quantities include the bend, splay, and twist scalars that appear in the Frank free energy density \cite{GP95}. As these intrinsic quantities are obtained from derivatives of the local director orientation, the values they assume are not independent of each other. The relations restricting the values of these intrinsic fields are called their compatibility conditions.}

Bent-core liquid crystals locally favor a phase of constant positive bend and vanishing splay and twist \cite{NE18,S18}. Such a phase was shown to be incompatible with the geometry of $\mathbb{R}^3$ \cite{Meyer1976}. In place of the incompatible locally favored phase, bent-core liquid crystals assume the ``closest" compatible phase, where the distance is measured using the system free energy {\cite{selingerAnnRev21}}. Such ``close-by" compatible phases include the twist-bend (heliconical) phase \cite{C+14} and the splay-bend phase \cite{CK19}, but also include many other less uniform (and less known) states. Identifying these states requires a comprehensive characterization of the collection of compatible states. Moreover, a clear characterization of the collection of compatible states in terms of local variables would advance our ability to solve inverse design problems where one seeks the local distortion fields that give rise to a specific configuration \cite{GAE19}.   

For two dimensions, it was shown that the compatibility condition consists of a single first-order partial differential equation in the splay and bend of the director field. It was also shown that when the bend and splay fields are compatible, they suffice to uniquely define a director field up to rigid motions \cite{NE18}. This result relied on the ability to construct a two-dimensional orthogonal coordinate system whose parametric curves are everywhere locally tangent to the director and the director normal. This construction greatly simplifies the geometric treatment, yet it can not be generalized to three dimensions. Consequently, for three-dimensional director fields, we are required to formulate the problem in terms of the spatial variation of a local orthogonal triad, an approach made formal through Cartan's method of moving frames \cite{dSE21,PA21}. While this approach proved very potent and concise, it is less accessible than other approaches (such as vector calculus). In what follows, we repeat the derivation of the compatibility conditions carried out in \cite{dSE21,PA21} using vector calculus  (as carried out in \cite{V19} for the special case of uniform phases). { We then} explore possible solutions to the general systems of equations in selected cases exhibiting some {simplifying symmetries}. 

The remaining of this manuscript is divided as follows. In Section \ref{sec:GradN}, we present the fundamental equations defining the local representation of the director gradients and discuss the relation of this approach to Cartan's method of moving frames. In Section \ref{sec::CompatibilityEqs}, we derive the corresponding compatibility conditions, which then provide {necessary conditions} for the existence of a director field with a prescribed set of deformation modes. {(Interpreting in which way the compatibility equations can be seen as sufficient conditions requires a formalism that would take us beyond the scope of this work, i.e., vector calculus tools. We briefly discuss how to achieve this goal in Appendix \ref{AppendixCompEqsAsSufficient}.)} In Section \ref{sec::SelectedCompatiblePhases}, we find specific solutions to these equations under the simplifying assumptions of additional symmetries. In other words, we provide several examples of how to {use the compatibility conditions to} find compatible deformation modes. Finally, in the concluding remarks, Section \ref{sec::Conclusion}, we present alternative ways of interpreting the compatibility conditions.

\section{Local representation of the director gradient}
\label{sec:GradN}

{Consider a liquid-crystalline phase in a domain $U\subset\mathbb{R}^3$ described by a unit vector field $\hatn$ with Cartesian components $n_j$. We shall refer to $\hatn$ as the director field. Let $\hatp$ and $\hatq$ be two orthogonal unit vectors that span the space normal to $\hatn$, and $p_{j}$ and $q_{j}$ be their corresponding Cartesian components.\footnote{{As a concrete example, if $\hatn$ does not coincide with a coordinate direction, say the $x$-direction $\hat{x}$ in the domain, we can set $\hatp=\hatn\times\hat{x}/\Vert\hatn\times\hat{x}\Vert$ and $\hatq=\hatn\times\hatp$. However, it may often be useful to choose $\hatp$ and $\hatq$ with some extra properties that make them more appropriate to the context.}}} Following Machon and Alexander \cite{MA16} {and Selinger \cite{S18}}, we  denote (sum on repeated indices)
\begin{equation}\label{eq::MachonAlexDecompositionOfJij}
J^{n}_{ij}\equiv \partial_{i} n_{j}= -n_i(b_{p}p_{j}+b_{q}q_{j})+\frac{s}{2}(\delta_{ij}-n_in_j)+\frac{t}{2}\epsilon_{ijk}n_k+\Delta_{ij}, 
\end{equation}
where $\partial_{i}=\frac{\partial }{\partial x^{i}}$ and $x^{i}$ is the $i^{th}$ Cartesian coordinate ($i=\{1,2,3\}$) { and the deformation modes appearing in $J^n$
 are the components of the bend vector $\mathbf{b}=b_p\hatp+b_q\hatq=\hatn\times\nabla\times\hatn=-\hatn\cdot\nabla\hatn$ whose modulus  gives the bend $b=\Vert\mathbf{b}\Vert$, the splay $s=\nabla\cdot\hatn$, the twist $t=\hatn\cdot\nabla\times\hatn$, and the Cartesian components of the so-called biaxial splay $\Delta_{ij}$. (See Ref. \cite{S18} for a further discussion concerning the interpretation of the deformation modes.)}

The biaxial splay $\Delta_{ij}$ corresponds to the traceless and symmetric part of the gradient of $\hatn$.  Only recently has the biaxial splay been studied as a local property of the constituents of a liquid crystal, similar to the local bend, splay, and twist \cite{selingerPRE22}.  When it is expressed in the basis $\mathcal{F}=\{\hatn,\hatp,\hatq\}$, the biaxial splay takes the form
\[
(\Delta_{ij})_{\mathcal{F}} = \left(
\begin{array}{ccc}
    0 & 0 & 0  \\
    0 & \Delta_1 & \Delta_2 \\ 
    0 & \Delta_2 & -\Delta_1\\
\end{array}
\right),\, \text{i.e.,} \quad 
\Delta_{ij}=\Delta_1 p_i p_j+\Delta_2 (p_i q_j+q_i p_j)-\Delta_1 q_i q_j.
\]
The variation of the components of $\hatn$ above is given in terms of the components of $\hatp$ and $\hatq$. Thus, to obtain a complete system of equations, we need also to provide $J_{ij}^p=\partial_ip_j$ and $J_{ij}^q=\partial_iq_j$. The restriction that $\hatn,\hatp,\hatq$ are triply orthogonal unit vector fields allows us to express these gradients using only three additional scalar fields. We  thus obtain a closed system of three first-order equations prescribing the spatial variation of the triad $\hatn$, $\hatp$, and $\hatq$:
\begin{eqnarray}
\partial_i n_j & = & \left[-n_ib_p+\left(\frac{s}{2}+\Delta_1\right)p_i-\left(\frac{t}{2}-\Delta_2\right)q_i\right]\,p_j \nonumber \\
& & +\left[-n_ib_q+\left(\frac{t}{2}+\Delta_2\right)p_i+\left(\frac{s}{2}-\Delta_1\right)q_i\right]\,q_j, \label{eq::deliNj} \\ 
\partial_i p_j & = & \left[b_pn_i-\left(\frac{s}{2}+\Delta_1\right)p_i-\left(-\frac{t}{2}+\Delta_2\right)q_i\right]n_j+[\alpha n_i+\beta p_i+\gamma q_i]q_j, \label{eq::deliPj} \\ 
\partial_i q_j & = &  \left[b_qn_i-\left(\frac{t}{2}+\Delta_2\right)p_i-\left(\frac{s}{2}-\Delta_1\right)q_i\right]n_j+[-\alpha n_i-\beta p_i-\gamma q_i]p_j. \label{eq::deliQj}
\end{eqnarray}

These are the fundamental reconstruction equations for a director field in $\mathbb{R}^3$, and they form the basis for all the following calculations. We may alternatively express $J^n$, $J^p$, and $J^q$ in matrix form through
\begin{equation}\label{eq::JInBasisNPQ}
J^n =R^T M^n
R,\,\quad
J^p =R^T M^p
R,\quad\mbox{and}\quad
J^q =R^T M^q
R,\,\quad
\end{equation}
where
\begin{equation}\nonumber
R=\left(
    \begin{array}{ccc}
        n_1 & n_2 & n_3  \\
        p_1 & p_2 & p_3  \\
        q_1 & q_2 & q_3  \\
    \end{array}
    \right),\quad
    M^n=\left(
    \begin{array}{ccc}
        0 & -b_p                  & -b_q \\ [6pt]
        0 &  \dfrac{s}{2}+\Delta_1 & \dfrac{t}{2}+\Delta_2 \\ [6pt]
        0 & -\dfrac{t}{2}+\Delta_2 & \dfrac{s}{2}-\Delta_1 \\
    \end{array}
    \right),
\end{equation}
\begin{equation}\label{eq::MInBasisNPQ}
M^p=\left(
    \begin{array}{ccc}
        b_p                  & 0 & \alpha \\ [6pt]
         -\dfrac{s}{2}-\Delta_1 & 0 & \beta \\ [6pt]
        \dfrac{t}{2}-\Delta_2 & 0 & \gamma \\
    \end{array}
    \right),\,\mbox{ and }\quad
M^q=\left(
    \begin{array}{ccc}
        b_q                  & -\alpha & 0 \\ [6pt]
        -\dfrac{t}{2}-\Delta_2 & -\beta & 0 \\ [6pt]
        -\dfrac{s}{2}+\Delta_1 & -\gamma & 0 \\
    \end{array}
    \right).
\end{equation}

\begin{remark}[On the relation to Cartan's method of moving frames]
For any given vectors $\mathbf{v}$ and $\mathbf{w}$, we have $(\mathbf{v}\cdot\nabla\hatn)\cdot\mathbf{w}=v_iJ_{ij}^nw_j$. Comparison with the approach via the moving frame method \cite{dSE21} shows that $(\mathbf{n}_i\cdot\nabla\mathbf{n}_k)\cdot\mathbf{n}_j=\eta_k^j(\mathbf{n}_i)$ and, therefore, the second and third columns of $M^n$ are respectively given by $(\,\eta_1^2(\hatn)\quad\eta_1^2(\hatp)\quad\eta_1^2(\hatq)\,)^T$ and $(\,\eta_1^3(\hatn)\quad\eta_1^3(\hatp)\quad\eta_1^3(\hatq)\,)^T$. Analogously, the first and third columns of $M^p$ are respectively given by $(\,\eta_2^1(\hatn)\quad\eta_2^1(\hatp)\quad\eta_2^1(\hatq)\,)^T$ and $(\,\eta_2^3(\hatn)\quad\eta_2^3(\hatp)\quad\eta_2^3(\hatq)\,)^T$, while the first and second columns of $M^q$ are respectively given by $(\,\eta_3^1(\hatn)\quad\eta_3^1(\hatp)\quad\eta_3^1(\hatq)\,)^T$ and $(\,\eta_3^2(\hatn)\quad\eta_3^2(\hatp)\quad\eta_3^2(\hatq)\,)^T$. (Note that $\eta_i^j=-\eta_j^i$, which implies that $M^n$ and $M^p$, or $M^n$ and $M^q$, or $M^p$ and $M^q$, share one similar column up to a sign.) {Under this identification, the compatibility conditions derived in the next section as the solvability condition for the  set of first-order partial differential equations \eqref{eq::deliNj}, \eqref{eq::deliPj}, and \eqref{eq::deliQj}, coincide with the structure equations of the corresponding moving frame.} 
\end{remark}

\section{Compatibility equations}
\label{sec::CompatibilityEqs}

One of the main challenges in constructing the compatibility conditions that prescribe the necessary relations between the intrinsic local fields describing a director in $\mathbb{R}^3$ is that it is \textit{a priori} unknown how many such fields are required to describe such a director uniquely. The fundamental equations above, Eqs. \eqref{eq::deliNj}, \eqref{eq::deliPj}, and \eqref{eq::deliQj},  make use of nine scalar (and pseudoscalar) local intrinsic quantities, thus bounding the number of the intrinsic descriptors required to define the texture at some neighborhood uniquely. In the next section, we will show that the number of intrinsic fields necessary to fully describe a director field in $\mathbb{R}^3$ may be further reduced to only five. However, for simplicity and properly constructing the compatibility conditions, we first assume all the intrinsic fields that appear in Eqs. \eqref{eq::deliNj}, \eqref{eq::deliPj}, and \eqref{eq::deliQj} are known, as are the components of their derivatives along the vectors $\hatn$, $\hatp$, and $\hatq$. We then consider Eqs. \eqref{eq::deliNj}, \eqref{eq::deliPj}, and \eqref{eq::deliQj} as partial differential equations (PDEs) describing the spatial evolution of the orthonormal triplet $\hatn$, $\hatp$, and $\hatq$. Given some initial value for the orthonormal triplet at a point in a domain (as well as all the fields that appear in the PDEs), we could obtain the values the triplet obtains in its neighborhood by integrating
\[
\nabla\hatn=J^n,\quad \nabla\hatp=J^p,\quad \text{and}\quad
\nabla\hatq=J^q.
\]
Such a system of PDEs is meaningful and could be solved only if its solvability conditions are satisfied. In the present case, these necessary and sufficient solvability conditions on a simply connected domain read {(see Theorem 10.9 of \cite{Apostol}, or Sect. 8 of Chapter 2 of \cite{LeviCivita})}:
\begin{equation}\label{eq::CurlsOfGradientsVanish}
    \nabla\times\nabla n_j = 0,\quad \nabla\times\nabla p_j = 0,\quad\mbox{and}\quad\nabla\times\nabla q_j = 0,\quad j\in\{1,2,3\}.
\end{equation}

Had all the components of the gradients $J^n$, $J^p$, and $J^q$ been independent, these equations would have formed 27 differential equations in 27 unknown fields. However, the expressions above teach us that the gradients may be fully expressed in only nine fields: $b_p,b_q,s,t,\Delta_1,\Delta_2,\alpha,\beta,\gamma$. Moreover, because $\hatn,\hatp$, and $\hatq$ are unit vectors, three of the nine components vanish identically in each of their curl equations. Due to the mutual orthogonality of these vectors, only nine of the remaining eighteen equations are independent. Thus, the compatibility conditions of the system consist of nine equations in nine fields. {Before we derive these equations in more easily interpretable physical terms, it is essential to emphasize that different representations of the same differential system, while algebraically equivalent to one another, may carry different information as they pertain to different unknowns and prescribed data. Consequently, necessary and sufficient conditions, such as Eq. \eqref{eq::CurlsOfGradientsVanish}, may prove to be only necessary but not sufficient once reformulated. 
%For example, treating the compatibility equations as sufficient conditions requires the formalism of differential forms instead of vector calculus. (See Appendix \ref{AppendixCompEqsAsSufficient}.)
% this is not an example, and we already stated this.
} 

We begin the reformulation by defining the commutation tensor $C^n_{k\ell}$:
\begin{equation}
    C^n_{k\ell} = \epsilon_{ijk}\partial_i \partial_jn_{\ell}= \epsilon_{ijk}\partial_i J_{j\ell} .
\end{equation}
Note that $\nabla\times\nabla n_{\ell}=(C^n_{1\ell},C^n_{2\ell},C^n_{3\ell})$ must vanish in Euclidean space. Analogously, we can define $C_{k\ell}^p = \epsilon_{ijk}\partial_i J_{j\ell}^p$ and $C_{k\ell}^q = \epsilon_{ijk}\partial_i J_{j\ell}^q$. The compatibility conditions require all the components of the three commutation tensors to vanish. However, contracting the obtained commutation tensors with the vectors of the frame $\{\hatn,\hatp,\hatq\}$ yields more concise and transparent equations of the form: $C^n_{k\ell}\,n_kn_{\ell}=0$, $C^n_{k\ell}\,n_kp_{\ell}=0$, etc.

The contractions with $C^n_{k\ell}$ will provide six non-trivial equations, Eqs. \eqref{eq::CompCondUsingABCR1212}--\eqref{eq::CompCondUsingABCR1323} at the end of this section, while the contractions with $C^p_{k\ell}$ will provide three additional non-trivial equations, Eqs. \eqref{eq::CompCondUsingABCR2312}--\eqref{eq::CompCondUsingABCR2323} at the end of this section. The resulting nine first-order equations in nine fields could be further simplified. We use three of the equations to express $\alpha$, $\beta$, and $\gamma$, which describe the deformation modes of $\hatp$ and $\hatq$, in terms of $\{b_p,b_q,t,s,\Delta_1,\Delta_2\}$ and their gradients. Substituting these expressions in the compatibility equations yields \emph{six} compatibility conditions in terms of \emph{six} deformation modes of $\hatn$: $\{b_p,b_q,t,s,\Delta_1,\Delta_2\}$, three equations of the first order and three of second order.

There are six deformation modes in the expression for $\nabla\hatn$, Eq. \eqref{eq::deliNj}. However, there exists gauge freedom in the choice of $\hatp$ and $\hatq$, which implies we need only five scalar fields. Indeed, we may choose a frame $\{\hatn,\hatp,\hatq\}$ where:
\begin{enumerate}[(i)]
    \item $\hatp=\frac{1}{b}\mathbf{b}$, which implies $b_q=0$. Geometrically, $\hatp$ and $\hatq$ have the same direction as the principal normal and binormal vector fields of the integral lines of the director $\hatn$; or
    \item $\hatp$ and $\hatq$ are the eigenvectors of the biaxial splay, which implies $\Delta_2=0$.
\end{enumerate}
Therefore, taking the gauge freedom into account, we may say the compatibility conditions consist of \emph{six} equations in \emph{five} deformation modes. The gauge invariant physical deformation modes can be taken to be the splay, $s$, bend (magnitude), $b$, twist, $t$, biaxial splay (magnitude), $\Delta=\sqrt{(\Delta_1)^2+(\Delta_2)^2}$, and the relative angle between the principal direction of the biaxial splay and the bend vector, $\phi$, which satisfies   {$\mathbf{b}\cdot\mathcal{D}\mathbf{b}=b^2\Delta\cos(2\phi)$, where $\mathcal{D}$ denotes the biaxial splay acting as an operator on the plane normal to $\hatn$}. 

\begin{remark}\label{remark::AlphaZero}
It may be useful to employ other gauge choices for $\hatp$ and $\hatq$. For example, $\hatp$ and $\hatq$ may be chosen to minimize rotation along the integral lines of the director. This choice implies that $\alpha=0$ \cite{B75}. (The equations of motion of the frame $\{\hatn,\hatp,\hatq\}$ along the integral curves of the director are given by the first set of equations in \eqref{eq::EvolEqsFrameNPQandPQN}.)
\end{remark}

\subsection{Obtaining the compatibility equations}

From Eq. \eqref{eq::deliNj}, the vector $\nabla\times\nabla n_{\ell}=(C^n_{1\ell},C^n_{2\ell},C^n _{3\ell})$ takes the form
\begin{eqnarray}
    \nabla\times\nabla n_{\ell} & = & {\nabla p_{\ell}\times[-b_p\hatn+(\frac{s}{2}+\Delta_1)\hatp+(-\frac{t}{2}+\Delta_2)\hatq]}\nonumber\\
    & &+ {\nabla q_{\ell}\times[-b_q\hatn+(\frac{t}{2}+\Delta_2)\hatp+(\frac{s}{2}-\Delta_1)\hatq]}\nonumber\\
    & &+ p_{\ell}\nabla\times[-b_p\hatn+(\frac{s}{2}+\Delta_1)\hatp+(-\frac{t}{2}+\Delta_2)\hatq]\nonumber\\
    & &+ q_{\ell}\nabla\times[-b_q\hatn+(\frac{t}{2}+\Delta_2)\hatp+(\frac{s}{2}-\Delta_1)\hatq].
\end{eqnarray}

As it is easier to compute dot products than to compute cross products, we are going to exploit the vector calculus identity
$$\nabla\times(\mathbf{A}\times\mathbf{B})=(\nabla\cdot\mathbf{B})\mathbf{A}-(\nabla\cdot\mathbf{A})\mathbf{B}+\mathbf{B}\cdot\nabla\mathbf{A}-\mathbf{A}\cdot\nabla\mathbf{B}$$
to write
\begin{eqnarray}
\nabla\times\hatn &=& \nabla\times(\hatp\times\hatq) = (\nabla\cdot\hatq)\hatp - (\nabla\cdot\hatp)\hatq+\hatq\cdot\nabla\hatp-\hatp\cdot\nabla\hatq, \nonumber\\
\nabla\times\hatp &=& \nabla\times(\hatq\times\hatn) = (\nabla\cdot\hatn)\hatq - (\nabla\cdot\hatq)\hatn+\hatn\cdot\nabla\hatq-\hatq\cdot\nabla\hatn,\nonumber\\
\nabla\times\hatq &=& \nabla\times(\hatn\times\hatp) = (\nabla\cdot\hatp)\hatn - (\nabla\cdot\hatn)\hatp+\hatp\cdot\nabla\hatn-\hatn\cdot\nabla\hatp.\nonumber
\end{eqnarray}
Now, from Eqs. \eqref{eq::deliNj}, \eqref{eq::deliPj}, and \eqref{eq::deliQj} we have
\begin{equation}
\nabla\cdot\hatn = s,\quad \nabla\cdot\hatp = b_p+\gamma,\quad \nabla\cdot\hatq = b_q-\beta,
\end{equation}
\begin{equation}\label{eq::EvolEqsFrameNPQandPQN}
    %\left\{
    \begin{array}{l}
         \hatn\cdot\nabla\hatn = -b_p\,\hatp-b_q\,\hatq  \\ [3pt]
         \hatn\cdot\nabla\hatp = b_p\,\hatn + \alpha\, \hatq \\ [3pt]
         \hatn\cdot\nabla\hatq = b_q\,\hatn - \alpha\, \hatp \\
    \end{array}
    %\right.
    ,\quad
    %\left\{
    \begin{array}{l}
         \hatp\cdot\nabla\hatp = \beta\,\hatq-(\frac{s}{2}+\Delta_1)\,\hatn \\ [3pt]
         \hatp\cdot\nabla\hatq =  - \beta \,\hatp-(\frac{t}{2}+\Delta_2)\,\hatn \\ [3pt]
         \hatp\cdot\nabla\hatn = (\frac{s}{2}+\Delta_1)\,\hatp + (\frac{t}{2}+\Delta_2)\, \hatq \\
    \end{array}
    %\right.
    ,
\end{equation}
and
\begin{equation}\label{eq::EvolEqsFrameQNP}
    %\left\{
    \begin{array}{l}
         \hatq\cdot\nabla\hatq = -(\frac{s}{2}-\Delta_1)\,\hatn-\gamma\,\hatp \\ [3pt]
         \hatq\cdot\nabla\hatn =  (\frac{s}{2}-\Delta_1)\, \hatq -(\frac{t}{2}-\Delta_2)\,\hatp \\ [3pt]
         \hatq\cdot\nabla\hatp = \gamma\, \hatq -(-\frac{t}{2}+\Delta_2)\,\hatn \\ 
    \end{array}
    %\right.
    .
\end{equation}
Thus, we can express the curl of each vector field in the frame as
\begin{eqnarray}
    \nabla\times\hatn &=&  t\,\hatn+b_q\,\hatp-b_p\,\hatq,\\
    \nabla\times\hatp &=&  \beta\hatn + [(\frac{t}{2}-\Delta_2)-\alpha]\hatp + (\frac{s}{2}+\Delta_1)\hatq, \\
    \nabla\times\hatq &=&   \gamma\hatn -(\frac{s}{2}-\Delta_1)\,\hatp + [(\frac{t}{2}+\Delta_2)- \alpha]\, \hatq.
\end{eqnarray}
On the other hand, using the identity $$(\mathbf{A}\times\mathbf{B})\times\mathbf{C}=(\mathbf{A}\cdot\mathbf{C})\mathbf{B}-(\mathbf{B}\cdot\mathbf{C})\mathbf{A},$$
we have {(for any scalar function $f$)}
\begin{eqnarray}
\hatn\times\nabla f &=& (\hatp\times\hatq)\times\nabla f = (\hatp\cdot\nabla f)\hatq - (\hatq\cdot\nabla f)\hatp, \nonumber\\
\hatp\times\nabla f &=& (\hatq\times\hatn)\times\nabla f = (\hatq\cdot\nabla f)\hatn - (\hatn\cdot\nabla f)\hatq,\nonumber\\
\hatq\times\nabla f &=& (\hatn\times\hatp)\times\nabla f = (\hatn\cdot\nabla f)\hatp - (\hatp\cdot\nabla f)\hatn.\nonumber
\end{eqnarray}

We may now substitute for $\hatn\times\nabla p_{\ell}$, $\hatn\times\nabla q_{\ell}$, $\hatp\times\nabla p_{\ell}$, $\hatp\times\nabla q_{\ell}$, $\hatq\times\nabla p_{\ell}$, $\hatq\times\nabla q_{\ell}$ and $\nabla\times\hatn$, $\nabla\times\hatp$, $\nabla\times\hatq$ in the commutation tensor $C^n_{k\ell}$. Finally, we may compute the six contractions {$C^n_{k\ell}\,q_kp_{\ell}$, $C^n_{k\ell}\,p_kp_{\ell}$, $C^n_{k\ell}\,n_kp_{\ell}$, $C^n_{k\ell}\,q_kq_{\ell}$, $C^n_{k\ell}\,p_kq_{\ell}$, and $C^n_{k\ell}\,n_kq_{\ell}$} (which must all vanish in $\mathbb{R}^3$) in order to obtain the following compatibility equations:
\begin{eqnarray}
0 &=& -(\frac{s}{2}+\Delta_1)_{,n}-b_{p,p}-b_{p}^2-\frac{s^2}{4}+\frac{t^2}{4}-s\Delta_1-(\Delta)^2+2\alpha \Delta_2+\beta b_q, \label{eq::CompCondUsingABCR1212} \\
0 &=& -(-\frac{t}{2}+\Delta_2)_{,n}-b_{p,q}-b_pb_q-s(-\frac{t}{2}+\Delta_2)-2\alpha \Delta_1+\gamma b_q, \label{eq::CompCondUsingABCR1213} \\
0 &=& -(-\frac{t}{2}+\Delta_2)_{,p}+(\frac{s}{2}+\Delta_1)_{,q}+tb_p-2\beta \Delta_1-2\gamma\Delta_2, \label{eq::CompCondUsingABCR1223} \\
0 &=& -(\frac{t}{2}+\Delta_2)_{,n}-b_{q,p}-b_pb_q-s(\frac{t}{2}+\Delta_2)-2\alpha \Delta_1-\beta b_p, \label{eq::CompCondUsingABCR1312} \\
0 &=& -(\frac{s}{2}-\Delta_1)_{,n}-b_{q,q}-b_q^2-\frac{s^2}{4}+\frac{t^2}{4}+s\Delta_1-(\Delta)^2-2\alpha \Delta_2-\gamma b_p, \label{eq::CompCondUsingABCR1313} \\
0 &=& -(\frac{s}{2}-\Delta_1)_{,p}+(\frac{t}{2}+\Delta_2)_{,q}+tb_q-2\beta \Delta_2+2\gamma\Delta_1. \label{eq::CompCondUsingABCR1323}
\end{eqnarray}
Here, $f_{,n}$, $f_{,p}$, and $f_{,q}$ indicate the derivative of $f$ in the direction of $\hatn$, $\hatp$, and $\hatq$, respectively.

The contractions $C^n_{k\ell}\,n_kn_{\ell}$, {$C^n_{k\ell}\,p_kn_{\ell}$, and $C^n_{k\ell}\,q_kn_{\ell}$} vanish trivially and provide no further compatibility equations. It remains to compute $C_{k\ell}^p=\epsilon_{ijk}\partial_i J_{j\ell}^p$ and $C_{k\ell}^q=\epsilon_{ijk}\partial_i J_{j\ell}^q$. Proceeding similarly as we did for $C^n_{k\ell}$, it turns out that the computation of $C_{k\ell}^p$ results in three additional independent  equations: 
\begin{eqnarray}
0 &=& \alpha_{,p}-\beta_{,n}-(b_q+\beta )(\frac{s}{2}+\Delta_1)+(b_p-\gamma)(\frac{t}{2}+\Delta_2)+\alpha(b_p+\gamma) , \label{eq::CompCondUsingABCR2312} \\
0 &=& \alpha_{,q}-\gamma_{,n}+(b_p-\gamma){(\frac{s}{2}-\Delta_1)}-(b_q+\beta )(-\frac{t}{2}+\Delta_2)+\alpha(b_q -\beta ), \label{eq::CompCondUsingABCR2313} \\
0 &=& \beta_{,q}-\gamma_{,p}-\beta^2-\gamma^2-t\alpha -\frac{s^2}{4}-\frac{t^2}{4}+(\Delta)^2 \label{eq::CompCondUsingABCR2323} .
\end{eqnarray}

%\luiz{Because the unknowns vectors here are implicitly, then the satisfaction of \eqref{eq::CompCondUsingABCR1212}--\eqref{eq::CompCondUsingABCR2323}}

%\efi{Moreover, it could be that \eqref{eq::CompCondUsingABCR1212}--\eqref{eq::CompCondUsingABCR2323} hold, but Eqs. \eqref{eq::deliNj}, \eqref{eq::deliPj}, and \eqref{eq::deliQj}.}

{Equations \eqref{eq::CompCondUsingABCR1212}--\eqref{eq::CompCondUsingABCR2323} constitute the compatibility conditions that obstruct path-independent integration of Eqs. \eqref{eq::deliNj}, \eqref{eq::deliPj}, and \eqref{eq::deliQj} to obtain the unknowns $\hatn,\hatp$, and $\hatq$. However, as the equations explicitly contain the unknowns $\hatn,\hatp$, and $\hatq$, they will only be interpreted and exploited here as necessary conditions. Seeing them as sufficient conditions is a subtle task. If not properly interpreted, one could conceive a combination of distortion fields and an orthonormal triad that will satisfy Eqs. \eqref{eq::CompCondUsingABCR1212}--\eqref{eq::CompCondUsingABCR2323} but will fail to comply with Eqs. \eqref{eq::deliNj}, \eqref{eq::deliPj}, and \eqref{eq::deliQj}. (See Example \ref{exe::IncompatiblePlanePhase} in Appendix \ref{AppendixCompEqsAsSufficient}.) To see these equations as  sufficient conditions, one might need to formally invoke the cotangent bundle as a prescribed quantity, as carried out in \cite{T71}, and use the more powerful theory of exterior differential systems \cite{Bryant+91}. These remain outside the scope of the present work.} 

%To elevate the set of compatibility conditions to the degree of necessary and sufficient conditions, one may resort to more powerful tools in the formalism of differential forms and see \eqref{eq::CompCondUsingABCR1212}--\eqref{eq::CompCondUsingABCR2323} as an Exterior Differential System. In these cases, one must . See Appendix \ref{AppendixCompEqsAsSufficient} for more details.}
%Such a context, though, is beyond the scope of this work, i.e., vector calculus tools. We briefly discuss how this goal can be achieved in Appendix \ref{AppendixCompEqsAsSufficient}.}

Note that the variation of the director as expressed in $\nabla\hatn$ depends only on six deformation modes (or the five gauge invariant quantities). However, integration of the equations for $\nabla\hatn$ also requires knowledge of $\alpha$, $\beta$, and $\gamma$, which describe the spatial variation of $\hatp$ and $\hatq$. It is, therefore, natural to ask whether these variables are independent of the fields describing the deformation modes of $\hatn$ or could be eliminated from the equations. As we next show, the latter is indeed the case; as the first six compatibility conditions contain no derivatives of the fields $\alpha$, $\beta$, and $\gamma$, we may use three of these equations to express them using the six fields  $\{b_p,b_q,t,s,\Delta_1,\Delta_2\}$ and their first derivatives. When substituted back to the equations, we obtain six differential equations in the six unknown fields describing the deformations of $\hatn$;
three equations of the first order and three equations of the second order.

If $\Delta_1^2+\Delta_2^2\not=0$, then combining Eqs. \eqref{eq::CompCondUsingABCR1223} and \eqref{eq::CompCondUsingABCR1323} (multiplied by $\Delta_1$ and $\Delta_2$, respectively) allows us to express $\beta$ as
\begin{eqnarray}
    \beta & = & \frac{\Delta_1t_{,p}-\Delta_2s_{,p}+\Delta_1s_{,q}+\Delta_2t_{,q}}{4\Delta^2}\nonumber\\
    & &-\frac{\Delta_1\Delta_{2,p}-\Delta_2\Delta_{1,p}}{2\Delta^2}+\frac{(\Delta^2)_{,q}}{4\Delta^2}+t\,\frac{b_p\Delta_1+b_q\Delta_2}{2\Delta^2}.\label{eq::BetaAsFunctionOfDeformationModes}
\end{eqnarray}
Now, a different linear combination of these two equations (essentially multiplied by $\Delta_2$ and $-\Delta_1$, respectively) allows us to express $\gamma$ as
\begin{eqnarray}
    \gamma & = & \frac{\Delta_2t_{,p}+\Delta_1s_{,p}-\Delta_1t_{,q}+\Delta_2s_{,q}}{4\Delta^2}\nonumber\\
    & &+\frac{\Delta_2\Delta_{1,q}-\Delta_1\Delta_{2,q}}{2\Delta^2}-\frac{(\Delta^2)_{,p}}{4\Delta^2}+t\,\frac{b_{p}\Delta_2-b_{q}\Delta_1}{2\Delta^2}.\label{eq::GammaAsFunctionOfDeformationModes}
\end{eqnarray}
Finally,   {summing Eq.}  \eqref{eq::CompCondUsingABCR1212}   {multiplied by $\Delta_2$, Eq.} \eqref{eq::CompCondUsingABCR1213}   {multiplied by $-\Delta_1$, Eq.} \eqref{eq::CompCondUsingABCR1312}   {multiplied by $-\Delta_1$, and Eq.} \eqref{eq::CompCondUsingABCR1313}   {multiplied by $-\Delta_2$, and  using} the above expressions for $\beta$ and $\gamma$ allow us to express $\alpha$ as
\begin{eqnarray}
    \alpha & = & \frac{\Delta_2b_{p,p}-\Delta_2b_{q,q}-\Delta_1b_{q,p}-\Delta_1b_{p,q}}{4\Delta^2}-\frac{b_{q}\Delta_{1,p}-b_{p}\Delta_{2,p}+b_{p}\Delta_{1,q}+b_{q}\Delta_{2,q}}{8\Delta^2}\nonumber\\
    & &+\frac{\Delta_2\Delta_{1,n}-\Delta_1\Delta_{2,n}}{2\Delta^2}-{\frac{b_p(s_{,q}+t_{,p})+b_q(t_{,q}-s_{,p})}{16\Delta^2}}\nonumber\\
    & &-\frac{tb^2}{8\Delta^2}+\frac{(b_{p}^2-b_{q}^2)\Delta_2-2b_{p}b_{q}\Delta_1}{4\Delta^2}    .\label{eq::AlphaAsFunctionOfDeformationModes}
\end{eqnarray}

On the other hand, if the biaxial splay vanishes, $\Delta_1^2+\Delta_2^2=0$, and  $b_p^2+b_q^2\not=0$, then we may use Eq. \eqref{eq::CompCondUsingABCR1212} and Eq. \eqref{eq::CompCondUsingABCR1312} to express $\beta$ as
\begin{eqnarray}
    \beta & = & \frac{b_qs_{,n}-b_pt_{,n}}{2b^2}{+\frac{b_qb_{p,p}-b_pb_{q,p}}{b^2}}+\frac{(s^2-t^2)b_q-2stb_p}{4b^2}\label{eq::BetaAsFunctionOfDeformationModesDeltaZero}
\end{eqnarray}
and {we may use Eqs. \eqref{eq::CompCondUsingABCR1213} and \eqref{eq::CompCondUsingABCR1313} to} express $\gamma$ as
\begin{eqnarray}
     \gamma & = & -\frac{b_ps_{,n}+b_qt_{,n}}{2b^2}+\frac{b_qb_{p,q}-b_pb_{q,q}}{b^2}-\frac{(s^2-t^2)b_p+2stb_q}{4b^2}.\label{eq::GammaAsFunctionOfDeformationModesDeltaZero}
\end{eqnarray}
Finally, when the biaxial splay vanishes, we can write $\alpha$ as a function of the deformation modes by substituting for $\beta$ and $\gamma$ in Eq. \eqref{eq::CompCondUsingABCR2323} or we can get rid of $\alpha$ by choosing $\hatp$ and $\hatq$ such that $\alpha=0$   (see Remark \ref{remark::AlphaZero}).

\section{Selected compatible phases: specific solutions}
\label{sec::SelectedCompatiblePhases}

{Equations \eqref{eq::CompCondUsingABCR1212}-\eqref{eq::CompCondUsingABCR2323} form the compatibility conditions for the fundamental equations \eqref{eq::deliNj}, \eqref{eq::deliPj}, and \eqref{eq::deliQj}. The task of finding compatible phases has several approaches. The first and most systematic one is to reformulate \eqref{eq::CompCondUsingABCR1212}-\eqref{eq::CompCondUsingABCR2323} as an Exterior Differential System  for the unknown local intrinsic fields. However, this formalism is beyond the scope of this work. The second approach assumes the Cartesian components of the deformation fields' gradients are known, identifying the compatibility conditions as algebraic equations in the components of the triad $\{\hatn,\hatp,\hatq\}$ and solving for them directly. The solutions obtained, though, may not satisfy equations \eqref{eq::deliNj}, \eqref{eq::deliPj}, and \eqref{eq::deliQj}, and their satisfaction needs to be imposed. Thus, both approaches require the calculation of additional conditions.
%However, each approach requires calculating additional compatibility conditions, treating the local intrinsic fields as the unknowns. (See concluding remarks for further discussion.)
We, therefore, do not follow these approaches but a third, simpler one. We seek solutions to simplified versions of the compatibility conditions obtained by imposing restrictions on the values of the deformation modes, such as setting some of them to constants. Given enough constraints, the compatibility conditions simplify and can be interpreted, allowing us to characterize the resulting phases. In this method, explicit constructions guarantee the sought solutions' existence. Conversely, without explicit constructions, we cannot assure a phase with the sought-simplified properties exists.}

{We begin this section by presenting the recently studied phases where all the deformation modes are constant. We then discuss phases where only one deformation mode is not constant, and finally, we investigate director fields with only two non-vanishing deformation modes.}

\subsection{Directors with uniform distortion fields}
\label{subsect::UniformDirectors}

Among all director fields, the simplest ones are the so-called uniform distortion directors, i.e., those director fields with all deformation modes constant in space. In this case, it is known that \cite{V19} (see also \cite{dSE21}):
\begin{enumerate}[(a)]
    \item if $\Delta=0$, then $b=s=t=0$, i.e., there is only the trivial solution;
    \item if $\Delta\not=0$, then $s=0$, $t=\pm2\Delta$, and $\phi=\frac{(2k+1)\pi}{4}$, $k\in\{0,1,2,3\}$, where $\phi$ is again the angle formed by the bend vector and the principal direction of the biaxial splay. In addition, the bend vector $\mathbf{b}$ bisects the principal directions of the biaxial   {splay: if} $t=2\Delta$, then $k=0$ or $k=2$ and if $t=-2\Delta$ then $k=1,3$. 
\end{enumerate}

\begin{figure}[t]
    \centering
    \includegraphics[width=\linewidth]{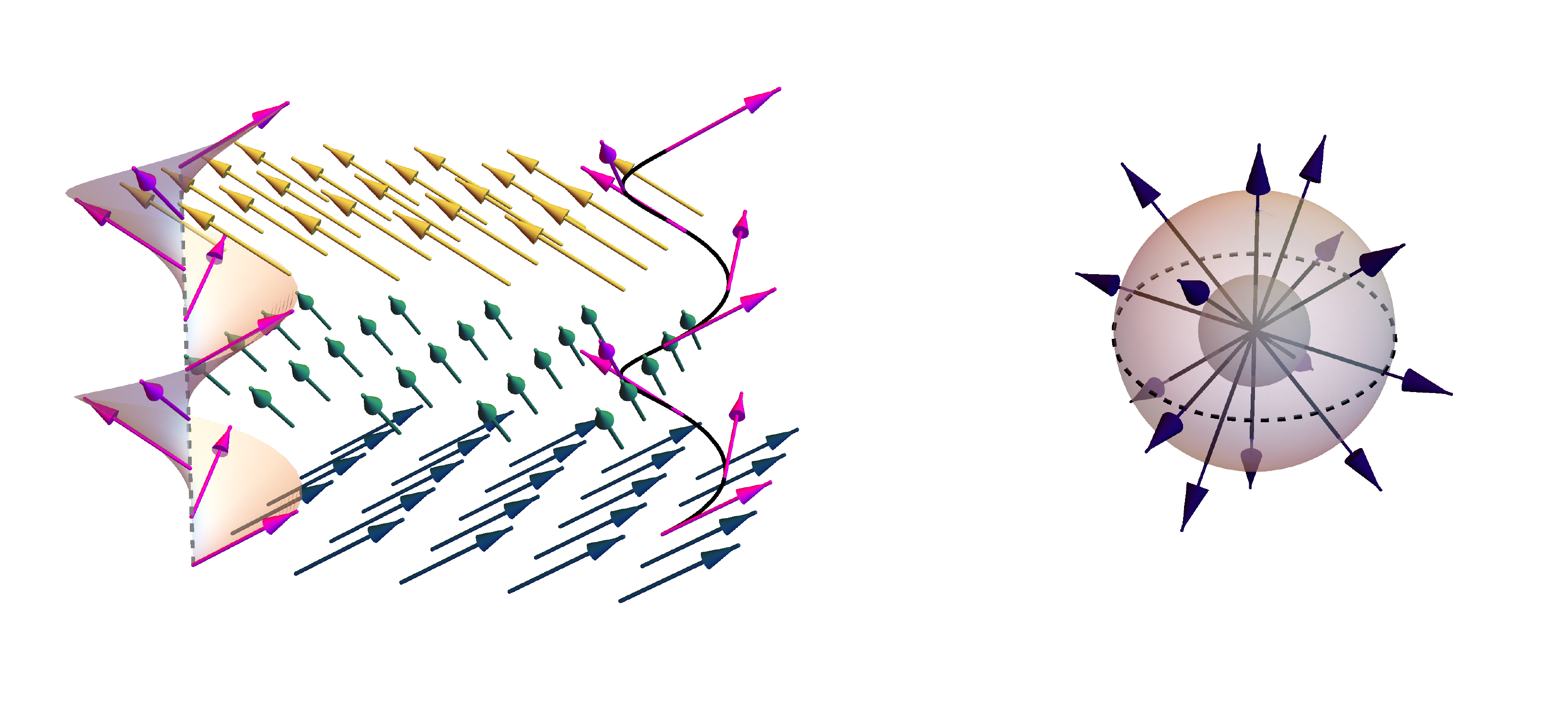}
    \caption{{Uniform distortion and splay-only phases. In the figures, the colors of the director vectors are just a guide to the eyes. (Left) The heliconical uniform distortion phase in which the director rotates at a constant rate along a fixed direction (not necessarily perpendicular to the director) while displaying translational invariance in the plane perpendicular to the screw axis \cite{V19}. The director is thus tangent to a helical conic surface from which it inherits its name. Integral curves of the director (solid black line on the right) provide a foliation of space by parallel helices of curvature $\kappa=b$ and torsion $\tau=-b^2/{2\Delta}$. (Right) A phase with all deformation modes equal to zero except for the splay. A splay-only phase is often termed a hedgehog phase. The integral lines of the hedgehog director can be described as the unit normal field of a family of concentric spheres.}    
    }\label{fig:HiliconicalndHedgehog}
\end{figure}

Geometrically, the two families of uniform phases correspond to a foliation of space by parallel helices (one family of solutions is the mirror image of the other), {see Fig. \ref{fig:HiliconicalndHedgehog}, Left}. The geometry of the helices, namely, their values of the curvature $\kappa$ and the torsion $\tau$, depends on the two free parameters, the biaxial splay $\Delta$ and the bend {$b$}:
\[
\kappa = b\quad\mbox{ and }\quad\tau=\alpha= \mp\frac{b^2}{4\Delta}+\frac{(b_{p}^2-b_{q}^2)\Delta_2-2b_{p}b_{q}\Delta_1}{4\Delta^2}.
\]
Remember that a helix of radius $r$ and pitch $p$, $\vartheta\mapsto(r\cos\vartheta,r\sin\vartheta,p\vartheta)$, has curvature $\kappa=\frac{r}{r^2+p^2}$ and torsion $\tau=\frac{p}{r^2+p^2}$. Conversely, a helix of curvature $\kappa$ and torsion $\tau$ has radius $r=\frac{\kappa}{\kappa^2+\tau^2}$ and pitch $p=\frac{\tau}{\kappa^2+\tau^2}$. 

\subsection{Directors with a single non-uniform distortion field}

We now investigate whether it is possible for a director field in Euclidean space to have all but one deformation mode constant. The theorems below show that this is not generally possible except for a few special cases, thus revealing an interesting rigidity of uniform distortion fields.

\begin{theorem}[Rigidity of non-uniform directors. I]\label{thrRigNonUnifDelta}
Let $\hatn$ be a director field in Euclidean space with deformation modes $b_p,b_q,s,t,\Delta_1=\Delta$, and $\Delta_2=0$, i.e., $\hatp$ and $\hatq$ are the {eigendirections} of the biaxial splay. If $b_p,b_q,s$, and $t$ are constant, then the modulus of the biaxial splay $\Delta$ must also be constant.
\end{theorem}

\begin{theorem}[Rigidity of non-uniform directors. II]
Let $\hatn$ be a director field in Euclidean space with deformation modes $b_p=b,b_q=0,s,t,\Delta_1$, and $\Delta_2$, i.e., $\hatp$ is the normalized bend vector. If $\Delta_{1},\Delta_{2},s$, and $t$ are constant, but $t\not=-2\Delta_1$ and $t\not=0$, then the bend $b$ must be also constant.
\label{thrRigNonUnifBend}
\end{theorem}

\begin{theorem}[Rigidity of non-uniform directors. III]
Let $\hatn$ be a director field in Euclidean space with deformation modes $b_p,b_q,t,s$ and vanishing biaxial splay $\Delta_1=\Delta_2=0$.
\begin{enumerate}[(a)]
    \item If $b_p,b_q$, and $t$ are constant and $b_p^2+b_q^2$ does not vanish, then the splay $s$ must be also constant.
    \item If $b_p,b_q$, and $s$ are constant, then the twist $t$ must be also constant.
\end{enumerate}
\label{thrRigNonUnifTwistOrSplay}
\end{theorem}

The proofs of Theorems  \ref{thrRigNonUnifDelta}, \ref{thrRigNonUnifBend}, and \ref{thrRigNonUnifTwistOrSplay} can be found in the Appendices \ref{appendixProofThm1}, \ref{appendixProofThm2}, and \ref{appendixProofThm3}, respectively. 

\subsubsection{Non-uniform splay phases}

Let us understand the exceptions to Theorems \ref{thrRigNonUnifDelta}, \ref{thrRigNonUnifBend}, and \ref{thrRigNonUnifTwistOrSplay}. The first possibility for a single non-uniform deformation mode director field corresponds to the case where $b=0,t=0,\Delta=0$, but $s$ is non-constant. An example of such deformation mode is the hedgehog director field, {see Fig. \ref{fig:HiliconicalndHedgehog}, Right}. Geometrically, the director $\hatn$ corresponds to the normals of a family of concentric spheres. The value of the splay at a point located at a sphere of radius $R$ is $s=\pm2/R$. It turns out that this is the only non-trivial solution. To see that, we first need to take into account the following geometric facts:
\begin{enumerate}[(a)]
    \item if $t=0$, then there exists a foliation of {(a possibly finite portion of)} space by surfaces such that the unit normal of each leaf is given by the director $\hatn$ \cite{Aminov}, {or Chap. 3 of Ref. \cite{DoCarmo94}, Exercise 12};
    \item if $t=0$, then the shape operator $A$ of each leaf of the foliation in (a) is given by (see Subsect. \ref{subsectZeroTwist}) 
    $$A=-\rmd\hatn=-\frac{s}{2}I-\left(\begin{array}{cc}
        \Delta_1 & \Delta_2  \\
        \Delta_2 &-\Delta_1\\
    \end{array}\right),$$
    where $I$ is the {identity} operator acting on the plane orthogonal to $\hatn$;
    \item if $t=0$ and $b=0$, then the foliation in (a) is given by a family of parallel surfaces \cite{Aminov}. In this case, we can parametrize the region of $\mathbb{R}^3$ where $\hatn$ is defined as
    \[
    \mathbf{R}(u,v,w) = \mathbf{r}(u,v)+w\hatn(u,v).
    \]
    Each surface $w=\mbox{const.}$ is parallel to a prescribed surface $\Sigma^2$ parametrized by $(u,v)\mapsto\mathbf{r}(u,v)$.
\end{enumerate}

Now, assume a phase exists with $t=0$, $b=0$, and $\Delta=0$. As $\hatn$ has $t=0$ and $b=0$, then $\hatn$ is the field of unit normals of a family of parallel surfaces. Since, in addition,  $\Delta=0$, then the shape operator of each leaf is just $A=-\frac{s}{2}I$, which implies the leaves are totally umbilical surfaces (planes or spheres). Therefore, $\hatn$ is either the trivial solution (foliation by parallel planes) or a hedgehog (foliation by concentric spheres, {see Fig. \ref{fig:HiliconicalndHedgehog}, Right.}).

\subsubsection{Non-uniform bend phases}
\label{subsubsect::NonUnifBend}

\begin{figure}[t]
    \centering
    \includegraphics[width=\linewidth]{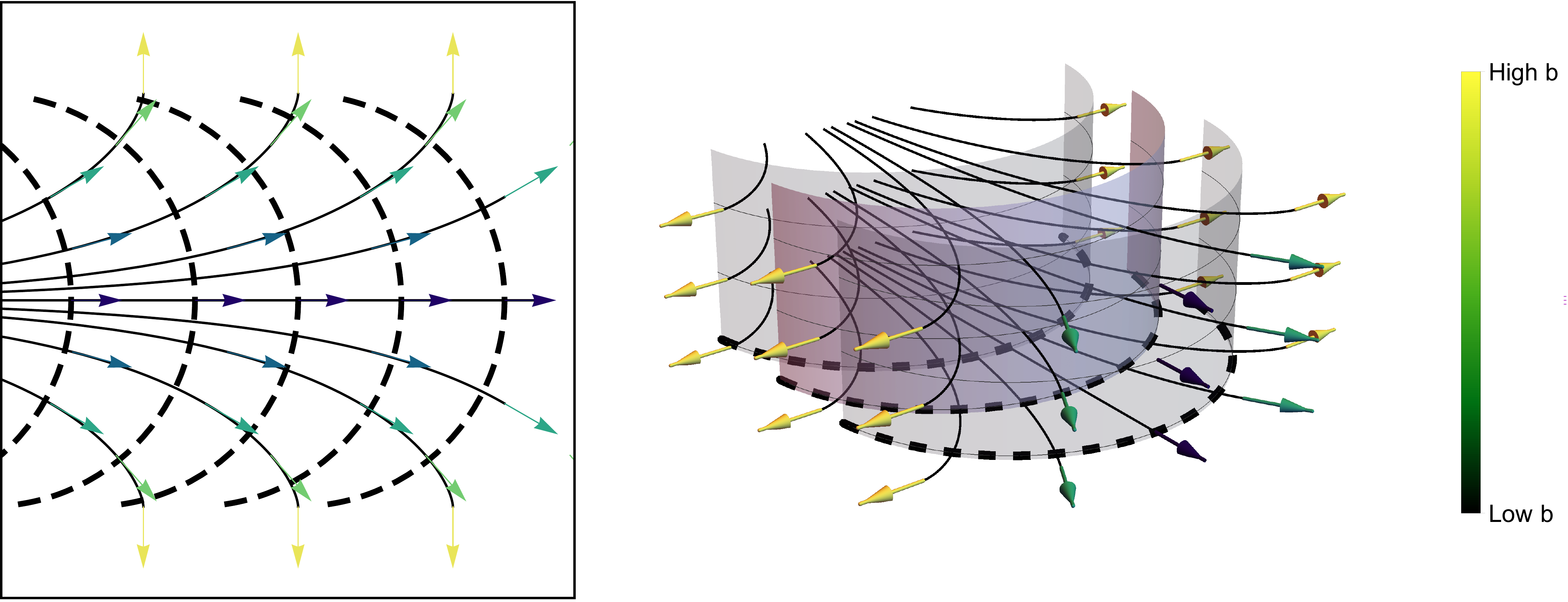}
    \caption{{Example of a non-uniform bend phase with $t=0$ and $s,\Delta$ constant. In the figures, the arrows representing the director are colored by the value of the bend $b$. (Left) The non-uniform bend and uniform splay $2d$ phase obtained by considering $\hatn$ orthogonal to the one-parameter family of circles of radius $\rho$ constant given by $\lambda\mapsto \mathbf{r}_{\lambda}(\theta)=(\lambda+\rho\cos\theta,\rho\sin\theta)$ (dashed  lines). The integral lines of $\hatn$ (full lines) have a bend equal to $b=\frac{1}{\rho}\tan\theta$ while the splay {$s=1/\rho$} is constant. (Right) The $3d$ phase obtained by translating the $2d$ phase on the left in the direction orthogonal to the plane of the $2d$ phase. The director $\hatn$ is orthogonal to a family of cylinders of the same radius and parallel axes but with distinct centers. Since $\hatn$ is normal to the foliation by the cylinders $\rho\mapsto\mathbf{R}_{\lambda}(\theta,z)=(\lambda+\rho\cos\theta,\rho\sin\theta,z)$, the twist vanishes $t=0$. In addition, as the leaves have uniform geometry, their shape operator $A_n=-d\hatn$ is constant, see Eq. \eqref{Def::FamilyShapeOperators}. Therefore, the biaxial splay and splay must be constant.  (The cylinders have distinct colors to ease the visualization.)}}
    \label{fig:3dHatNTzeroSDconstNon-UniformB}
\end{figure}

Let us now investigate the exceptions to Theorem \ref{thrRigNonUnifBend}. We may have a phase with non-uniform bend provided that $s,\Delta_1$, and $\Delta_2$ are constant, $\Delta_1^2+\Delta_2^2\not=0$, and  $t=0$ or $t=-2\Delta_2$.
\newline
\newline
\textit{a) Non-uniform bend phases with zero twist:} First assume that $t=0$ {and} $\Delta_1,\Delta_2$, and $s$ are constant, with $\Delta_1^2+\Delta_2^2\not=0$, and $b_q=0$, but $b_p=b$ non-uniform. We are going to show that we can construct $\hatn$ as the unit normal of a foliation of {a finite domain of} space by cylinders of the same radius and {with parallel axes, as illustrated in Fig.  \ref{fig:3dHatNTzeroSDconstNon-UniformB}.}

Here, the compatibility equations \eqref{eq::CompCondUsingABCR1212}--\eqref{eq::CompCondUsingABCR1323} become
\begin{eqnarray}
0 &=& -b_{,p}-b^2-\frac{s^2}{4}-s\Delta_1-(\Delta)^2+2\alpha \Delta_2,\label{eq::R1212DeltaTzeroSConstBnonConst} \\
0 &=& -b_{,q}-s\Delta_2-2\alpha\Delta_1,\label{eq::R1213DeltaTzeroSConstBnonConst} \\
0 &=& -2\beta \Delta_1-2\gamma \Delta_2,\label{eq::R1223DeltaTzeroSConstBnonConst} \\
0 &=& -s\Delta_2-2\alpha\Delta_1-\beta b,\label{eq::R1312DeltaTzeroSConstBnonConst}\\
0 &=& -\frac{s^2}{4}+s\Delta_1-(\Delta)^2-2\alpha\Delta_2-\gamma b,\label{eq::R1313DeltaTzeroSConstBnonConst}\\
0 &=& -2\beta \Delta_2+2\gamma \Delta_1. \label{eq::R1323DeltaTzeroSConstBnonConst}
\end{eqnarray}

From Eqs. \eqref{eq::R1223DeltaTzeroSConstBnonConst} and \eqref{eq::R1323DeltaTzeroSConstBnonConst}, we conclude that 
\begin{equation}
t,\nabla\Delta_1,\nabla\Delta_2,\nabla s,b_q=0\Rightarrow \beta,\gamma=0.    
\end{equation}
Now, adding Eq. \eqref{eq::R1312DeltaTzeroSConstBnonConst} multiplied by $\Delta_1$ and Eq. \eqref{eq::R1313DeltaTzeroSConstBnonConst} multiplied by $\Delta_2$, we can obtain for $\alpha$
\begin{equation}
t,\nabla\Delta_1,\nabla\Delta_2,\nabla s,b_q=0\Rightarrow   \alpha=-\frac{\Delta_2}{2\Delta^2}\left(\frac{s^2}{2}+\Delta^2\right),     
\end{equation}
which is constant. Let us exploit the remaining compatibility equations, Eq. \eqref{eq::CompCondUsingABCR2313}, which gives
\begin{equation}
    0 = b(\frac{s}{2}-\Delta_1)\Rightarrow s=2\Delta_1,
\end{equation}
where we are assuming $b\not=0$. (Otherwise, the phase would be uniform.) On the other hand, Eq. \eqref{eq::CompCondUsingABCR2323} gives
\begin{equation}
    0 = -\frac{s^2}{4}+\Delta_1^2+\Delta_2^2 \Rightarrow \Delta_2 = 0.
\end{equation}
  In particular, $\alpha=0$. As mentioned in the previous subsection, the condition $t=0$ implies that the director corresponds to the field of unit normals to leaves of a foliation by surfaces whose shape operators are given by $A=-\frac{s}{2}I-\mathcal{D}$, where $I$ is the identity operator and $\mathcal{D}$ is the traceless and symmetric operator acting on the plane orthogonal to $\hatn$ such that $\mathcal{D}_{11}=\Delta_1$ and $\mathcal{D}_{12}=\Delta_2$. From what we deduced so far, the director $\hatn$ corresponds to the field of unit normals of surfaces with shape operator
\begin{equation}
    A=-\rmd\hatn=-\frac{s}{2}I
    -\left(
    \begin{array}{cc}
        \Delta_1 & \Delta_2  \\
        \Delta_2 &-\Delta_1\\
    \end{array}
    \right)=\left(
    \begin{array}{cc}
        -s & 0  \\
        0 & 0 \\
    \end{array}
    \right).
\end{equation}
Therefore, the leaves of the foliation are cylinders of radius $R=1/s$ and axis $\hatq$. Substituting $\alpha,\beta,\gamma=0$, $s=2\Delta_1$, and $\Delta_2=0$ in Eq. \eqref{eq::deliQj} implies $\nabla\hatq=0$ and, therefore, it follows that the {cylinders' axes} are all parallel. 

Finally, substituting  $s=2\Delta_1$ in Eq. \eqref{eq::R1212DeltaTzeroSConstBnonConst} gives the following evolution equation for $b$ along $\hatp$
\begin{equation}\label{eq::bder2TzeroD1D2Sconst}
    -b_{,p} = b^2+4\Delta^2 = b^2+s^2.
\end{equation}
Noting that the system displays translational invariance along the direction of $\hatq$, Eq. \eqref{eq::bder2TzeroD1D2Sconst} can be interpreted as the compatibility condition for a $2d$ director with uniform splay and non-uniform bend \cite{NE18,NEcomment}.
\newline
\newline
\textit{b) Non-uniform bend phases with non-zero twist:} Assume that $\Delta_1,\Delta_2,s$ are constant, $t=-2\Delta_2\not=0$, and $b_q=0$ (so, $b_p=b$). We will show that $s=2\Delta_1$, which implies that $\hatn$ is a cholesteric with  pitch axis $\hatP=\hatq$. 

We say that $\hatn$ is a cholesteric with pitch axis $\hatP$ if $\hatP\cdot\nabla\hatn$ has no component along $\hatP$ \cite{B+14}. Following the notation in \cite{B+14}, we may define a new parameter $q$ from $\hatP\cdot\nabla\hatn$. The Eqs. \eqref{eq::EvolEqsFrameNPQandPQN} and \eqref{eq::EvolEqsFrameQNP} imply the director $\hatn$ is (i) a cholesteric with pitch axis $\hatP=\hatp$ if and only if $s=-2\Delta_1$ and (ii) a cholesteric with pitch axis $\hatP=\hatq$ if and only if $s=2\Delta_1$. A director field may have two, one, or no cholesteric pitch axes depending on whether the cholestericity $\mathcal{R}=t^2-2\sigma$ is positive, zero, or negative, respectively. Here, $\sigma$ denotes the saddle-splay \cite{B+14}, which can be expressed in terms of the local deformation modes by \cite{S18}:
\begin{equation}\label{Eq::SigmaAsFunctionOfSTD}
\sigma=\nabla\cdot[s\,\hatn-\hatn\cdot\nabla\hatn]=
\frac{1}{2}\left(s^2+t^2\right)-2\Delta^2.
\end{equation}

Using that our non-uniform bend phase satisfies $t=-2\Delta_2$ and $s=2\Delta_1$ implies that $\sigma=0$ and, therefore, the cholestericity is positive. Thus, there must exist a second cholesteric pitch. Indeed, in addition to $\hatP_1=\hatq$, the director $\hatn$ also has $\hatP_2=\frac{1}{\Delta^2}(\Delta_2\hatp-\Delta_1\hatq)$ as a cholesteric pitch axis. Note that neither $\hatP_1$ nor $\hatP_2$ provides a foliation (since $\hatP_i$ would be the field of unit normals, the director $\hatn$ would be tangent to the leaves): both pitch axes have twist $\hatP_1\cdot\nabla\times\hatP_1$ and $\hatP_2\cdot\nabla\times\hatP_2$ given by $-2\Delta_2(1+\frac{b^2}{4\Delta^2})$, where we used that $\alpha=\frac{\Delta_2}{2\Delta^2}b^2$ as {we are going to show in Eq. \eqref{alphaNon-uniformBend} below}.

In this case, the second pitch axis, $\hatP_2$, has an interesting property; its integral curves are straight lines. Indeed, this is the same as showing that the $\hatP_2$-integral curves have no curvature, i.e., $\hatP_2\cdot\nabla\hatP_2=0$. Using Eqs. \eqref{eq::EvolEqsFrameNPQandPQN} and \eqref{eq::EvolEqsFrameQNP} with $t=-2\Delta_2$ and $s=2\Delta_1$ constant, we have
\begin{eqnarray}
    \hatP_2\cdot\nabla\hatP_2 & = & \frac{1}{\Delta^2}\left[\Delta_2^2\,\hatp\cdot\nabla\hatp+\Delta_1^2\,\hatq\cdot\nabla\hatq-\Delta_1\Delta_2(\hatp\cdot\nabla\hatq+\hatq\cdot\nabla\hatp)\right]\nonumber\\
    & = & \frac{1}{\Delta^2}[\Delta_2^2(\beta\hatq-2\Delta_1\hatn)-\gamma\Delta_1^2\,\hatp-\Delta_1\Delta_2(- \beta \hatp+\gamma\hatq -2\Delta_2\hatn)] \nonumber\\
    & = & \frac{\beta\Delta_2-\gamma\Delta_1}{\Delta^2}(\Delta_1\,\hatp+\Delta_2\,\hatq).
\end{eqnarray}
Now, using Eq. \eqref{eq::AlphaBetaNon-uniformBendPhase}, to be proved below,  we obtain {$\beta\Delta_2-\gamma\Delta_1=0$} and, consequently, $\hatP_2\cdot\nabla\hatP_2=0$ as stated.

If we choose $\{\hatn,\hatP_2,\hatQ_2\equiv\hatn\times\hatP_2\}$ as a new frame, then the corresponding deformation modes $\{\tilde{b}_p,\tilde{b}_q,\tilde{t}=t,\tilde{s}=s,\tilde{\Delta}_1,\tilde{\Delta}_2,\tilde{\alpha},\tilde{\beta},\tilde{\gamma}\}$ are 
\begin{equation}
    \tilde{b}_p=\frac{b\Delta_2}{\Delta},\tilde{b}_q=\frac{b\Delta_1}{\Delta},\tilde{\Delta}_1=-\Delta_1,\tilde{\Delta}_2=\Delta_2,\tilde{\alpha}=\alpha,\tilde{\beta}=0,\tilde{\gamma}=-\frac{b\Delta_2}{\Delta}.
\end{equation}
Remember that $t$ and $s$ do not depend on the choice of frame.

Let us show that $s=2\Delta_1$, which will imply that $\hatn$ constitutes a cholesteric phase. Assuming that $\Delta_1,\Delta_2,s$ are constant, $t=-2\Delta_2\not=0$, and $b_q=0$,  the compatibility equations \eqref{eq::CompCondUsingABCR1212}--\eqref{eq::CompCondUsingABCR1323} become
\begin{eqnarray}
0 &=& -b_{,p}-b^2-\frac{s^2}{4}+\frac{t^2}{4}-s\Delta_1-(\Delta)^2+2\alpha \Delta_2,\label{eq::R1212DeltaTSConstBnonConst} \\
0 &=& -b_{,q}-s(-\frac{t}{2}+\Delta_2)-2\alpha\Delta_1,\label{eq::R1213DeltaTSConstBnonConst} \\
0 &=& tb-2\beta \Delta_1-2\gamma \Delta_2,\label{eq::R1223DeltaTSConstBnonConst} \\
0 &=& -s(\frac{t}{2}+\Delta_2)-2\alpha\Delta_1-\beta b,\label{eq::R1312DeltaTSConstBnonConst}\\
0 &=& -\frac{s^2}{4}+\frac{t^2}{4}+s\Delta_1-(\Delta)^2-2\alpha\Delta_2-\gamma b,\label{eq::R1313DeltaTSConstBnonConst}\\
0 &=& -2\beta \Delta_2+2\gamma \Delta_1. \label{eq::R1323DeltaTSConstBnonConst}
\end{eqnarray}
Since $\Delta_1^2+\Delta_2^2\not=0$, the coefficients $\beta $ and $\gamma $ are given by
\begin{equation}\label{eq::AlphaBetaNon-uniformBendPhase}
    \beta  = \frac{tb\Delta_1}{2\Delta^2}= -\frac{b\Delta_1\Delta_2}{\Delta^2}\,\mbox{ and }\,   \gamma  = \frac{tb\Delta_2}{2\Delta^2}= -\frac{b\Delta_2^2}{\Delta^2}.
\end{equation}
Then, Eq. \eqref{eq::R1312DeltaTSConstBnonConst} implies
\begin{equation}\label{alphaNon-uniformBend}
    \alpha = -\frac{\beta b}{2\Delta_1} = -\frac{t}{4\Delta^2}\,b^2 = \frac{\Delta_2}{2\Delta^2}\,b^2.
\end{equation}
Note that Eqs. \eqref{eq::CompCondUsingABCR2312} and \eqref{eq::CompCondUsingABCR2313} become
\begin{eqnarray}
0 &=& \alpha_{,p}-\beta_{,n}-\beta(\frac{s}{2}+\Delta_1)+\alpha(b+\gamma),\label{eq::R2312DeltaTSConstBnonConst} \\
0 &=& \alpha_{,q}-\gamma_{,n}+(b-\gamma)(\frac{s}{2}-\Delta_1)-2\beta\Delta_2-\alpha\beta.\label{eq::R2313DeltaTSConstBnonConst}
\end{eqnarray}
We can manipulate the compatibility equations and deduce that, see Eqs. \eqref{eq::bder2_TSD1D2Const} and \eqref{eq::bder3_TSD1D2Const} in the appendix, we can write
\begin{eqnarray}
    \Delta_2b_{,p}-\Delta_1b_{,q} & = & 2\Delta_2(\frac{t^2}{4}-\frac{s^2}{4}-\Delta^2)-b^2(\frac{t}{2}+\Delta_2)-st\Delta_1\nonumber\\
    & = & -2\Delta_2(\frac{s^2}{4}+\Delta_1^2)+2s\Delta_1\Delta_2=-2\Delta_2(\frac{s}{2}-\Delta_1)^2,
\end{eqnarray}
where in the second equality, we used that $t=-2\Delta_2$.

Finally, adding Eq. \eqref{eq::R2312DeltaTSConstBnonConst} multiplied by $-\Delta_2$ and Eq. \eqref{eq::R2313DeltaTSConstBnonConst} multiplied by $\Delta_1$, and using the relation above for $\Delta_2b_{,p}-\Delta_1b_{,q}$, give
\begin{equation}
    0 = \frac{4b\Delta_2^2}{\Delta_2}\left(\frac{s}{2}-\Delta_1\right)^2 \Rightarrow s = 2\Delta_1.
\end{equation}

\subsection{Vanishing twist phases}
\label{subsectZeroTwist}

It is known that the necessary and sufficient condition for the director $\hatn$ to be orthogonal to the leaves of a foliation of {(a possibly finite domain of)} space by surfaces is the vanishing of the twist $t=\hatn\cdot\nabla\times\hatn=0$ \cite{Aminov,DoCarmo94}. (As is often the case with if-and-only-if theorems, one of the implications is easy to prove. Indeed, if $\hatn=\nabla f/\Vert\nabla f\Vert$, i.e., $\hatn$ is a unit normal for the level sets of $f$, then performing the explicit calculations shows that $\hatn\cdot\nabla\times\hatn=0$.)

Let us investigate how to relate the leaves' geometry to the director's deformation modes. For a director $\hatn$ at a point $p$, the tangent plane of the leaf passing through $p$ is given by   $T_p=\{\mathbf{v}:\mathbf{v}\cdot\hatn=0\}$. 
We may then associate with $\hatn$ the family of shape operators that measure the variation of $\hatn$ along $T_p$
\begin{equation}\label{Def::FamilyShapeOperators}
    \begin{array}{rcccl}
        A_n & : & T_p & \to & T_p \\
            &  & \mathbf{v} &\mapsto & -(\mathbf{v}\cdot\nabla)\hatn .\\  
    \end{array}
    \quad
\end{equation}
Note that  $A_n(\mathbf{v})$ does belong to $T_p$: $\hatn\cdot\hatn=1\Rightarrow (\mathbf{v}\cdot\nabla\hatn)\cdot\hatn=0$.

Taking into account Eqs. \eqref{eq::EvolEqsFrameNPQandPQN} and \eqref{eq::EvolEqsFrameQNP}, the matrix coefficients of $A_n$ in the basis $\{\hatp,\hatq\}$ is precisely
\begin{equation}\label{Eq::MatrixFamilyShapeOperators}
    A_n = 
    \left(
    \begin{array}{cr}
       -(\hatp\cdot\nabla\hatn)\cdot\hatp  & -(\hatq\cdot\nabla\hatn)\cdot\hatp  \\
        -(\hatp\cdot\nabla\hatn)\cdot\hatq & -(\hatq\cdot\nabla\hatn)\cdot\hatq 
    \end{array}
    \right) 
    =
    \left(
    \begin{array}{cr}
       -(\frac{s}{2}+\Delta_1)  & -(-\frac{t}{2}+\Delta_2)  \\
        -(\frac{t}{2}+\Delta_2) & -(\frac{s}{2}-\Delta_1) 
    \end{array}
    \right).
\end{equation}
The operators $A_n$ are indeed the shape operators of the leaves of a foliation if, and only if, each $A_n$ is symmetric: $A_n=A_n^T\Leftrightarrow t=0$.

Note that $s=-\mbox{Tr}(A_n)$. In addition, using the relation between the saddle-splay, twist, splay, and biaxial splay given by Eq. \eqref{Eq::SigmaAsFunctionOfSTD}, we can write
\begin{equation}
    \sigma = 2\det(A_n).
\end{equation}
Thus, if we assume $t=0$, then the mean $H$ and Gaussian $K$ curvatures of the family of shape operators $A_n$ associated with the director $\hatn$ satisfy $s=-2H$ and $\sigma=2K$. Moreover, if $\{\hatp,\hatq\}$ are the eigenvectors of the biaxial splay, then they also correspond to the principal directions of each surface in the foliation orthogonal to $\hatn$.

\subsubsection{Zero bend and zero twist phases}

From Eq. \eqref{eq::EvolEqsFrameNPQandPQN}, we see that $\nabla\times\hatn=0$ if, and only if, $t=0$ and $b=0$. Then, we can locally write $\hatn=\nabla f$ for some scalar function $f$. Since the gradient of $f$ is of unit length, the length of an integral line of $\hatn$ connecting a point of the level set $\{f=c_1\}$ to a point of the level set $\{f=c_2\}$ is always $\vert c_2-c_2\vert$. It follows  that the distance between  the level sets $\{f=c_1\}$ and $\{f=c_2\}$ is precisely $\vert c_2-c_2\vert$. {An example of such a phase is given by the splay-only hedgehog phase, see Fig. \ref{fig:HiliconicalndHedgehog}, Right.}

We can state that $\hatn$ corresponds to the unit normals of a foliation by equidistant surfaces if, and only if, $t=0$ and $b=0$. In this case, we can parametrize the region where $\hatn$ is defined as
\[
(u,v,w)\mapsto\mathbf{R}(u,v,w) = \mathbf{r}(v,w)+u\,\hatn(v,w),
\]
where $\mathbf{r}$ parametrizes some chosen level set of $f$. As we will prove in Subsect. \ref{subsect::ZeroBend}, if $b=0$, the values of the deformations modes are then determined by the values they assume on the points of $\mathbf{r}(v,w)$. Using Eqs. \eqref{eq::SolutionSplayForBzero} and \eqref{eq::SolutionSadSplayForBzero} from Subsect. \ref{subsect::ZeroBend}, the mean and Gaussian curvatures of a surface of the foliation at a distance $u$ from $\mathbf{r}(v,w)$ are given by
\begin{equation}
    H = \frac{H_0-K_0u}{1-2H_0 u+K_0u^2}\quad\mbox{and}\quad K = \frac{K_0}{1-2H_0 u+K_0u^2},
\end{equation}
where $H_0$ and $K_0$ are the mean and Gaussian curvatures of $\mathbf{r}$. (See also \cite{DoCarmo76}, Exercise 11, Chapter 3.) 
 
\subsection{Zero splay and zero twist phases}

\begin{figure}
    \centering
    \includegraphics[width=\linewidth]{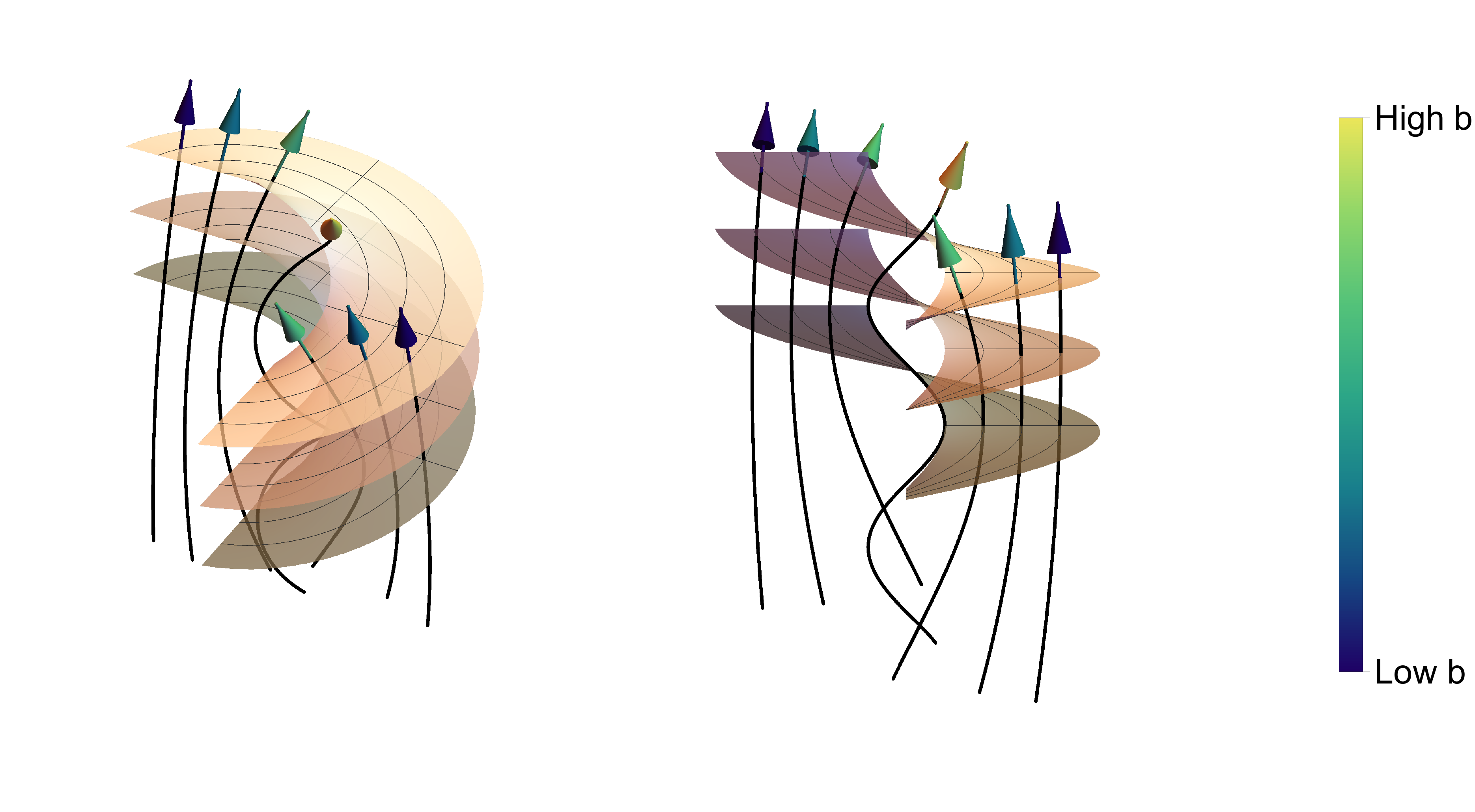}
    \caption{{Example of a zero twist and zero splay phase. In the figures, the arrows representing the director are colored by the value of the bend $b$. (Left and Right) Distinct views of the phase obtained from the unit normal vector field of a foliation of a portion of space by (vertically translated) helicoids, i.e., the levels sets of  $f(x,y,z)=z-p \arctan(y/x)$. The integral lines of the director field $\hatn=\nabla f/\Vert \nabla f\Vert$ are helices and they can be parametrized as $c(u)=(\rho_0\cos(-\omega\,u+\theta_0),\rho_0\sin(-\omega\,u+\theta_0),(\omega\rho_0^2/p)\,u+p\,\theta_0)$, where $\omega=(p/\rho_0)(\rho_0^2+p^2)^{-\frac{1}{2}}$ and the initial condition $c(0)$ is taken as a point of $f(x,y,z)=0$. The bend is the curvature of the helices: $b={p^2}/{\rho_0(p^2+\rho_0^2)}$. Consequently, the phase has a non-uniform bend. In addition, note that the closer to the common axis of the helicoids, i.e., the smaller the value of $\rho_0$, the higher the bend. (In the figures, the helicoids have distinct colors to ease the visualization.)}}
    \label{fig:FoliationByHelicoids}
\end{figure}
 
As the splay relates to the mean curvature of the leaves orthogonal to the director, $s=-2H$, we see that if $s=0$, then $\hatn$ is the unit normal of a foliation of {a portion of} space by minimal surfaces. 

{As an example, we may consider the level sets of the scalar function $f(x,y,z)=z-p\arctan(y/x)$. Each leaf of the foliation obtained from $f$ is a (vertically translated) helicoid of pitch $p$, which is a well-known minimal surface, see Fig. \ref{fig:FoliationByHelicoids}. The director is given by $\hatn=\nabla f/\Vert\nabla f\Vert$, where $\nabla f=(py/(x^2+y^2),-px/(x^2+y^2),1)$. Note that the director $\hatn$ and the gradient $\nabla f$ have the same integral lines up to reparametrization. A curve $c(\theta)=(x(\theta),y(\theta),z(\theta))$ is an integral line of $\nabla f$ if 
$$
\dot{x}(\theta)=\frac{p\,y(\theta)}{x(\theta)^2+y(\theta)^2}, \quad \dot{y}(\theta)=-\frac{p\,x(\theta)}{x(\theta)^2+y(\theta)^2}, \quad \mbox{and} \quad \dot{z}(\theta)=1.
$$
The solution $c(\theta)$ is a helix given by 
$$
c(\theta;\rho_0,\theta_0)=(\rho_0\cos(-\frac{1}{\rho_0^2}p\theta+\theta_0),\rho_0\sin(-\frac{1}{\rho_0^2}p\theta+\theta_0),\theta+p\theta_0),
$$
where the initial condition $c(0)$ is taken as a point on a chosen helicoid, e.g., $\Sigma_0:z=p\arctan\frac{y}{x}$.(\footnote{{The solution of $\dot{z}(\theta)=1$ is $z(\theta)=\theta+\zeta_0$. The fact that $c(0)$ must be a point of a helicoid implies $\zeta_0$ depends on $\theta_0$. If $c(0)\in\Sigma_0$, then $\zeta_0=p\theta_0$.}}) In this case, $\rho_0$ and $\theta_0$ parametrize $\Sigma_0$. Finally, parametrizing $c(\theta)$ by its arc length gives an integral line of the director. Note that $c(\theta)$ is a helix of radius $\rho_0$ and pitch $P=-\frac{\rho_0}{p}$. Thus, the integral lines of the director $\hatn$ have a handedness opposite to that of the helices foliating the family of helicoids orthogonal to $\hatn$. } 

\subsection{Zero biaxial splay and zero twist phases}

\begin{figure}[t]
    \centering
    \includegraphics[width=\linewidth]{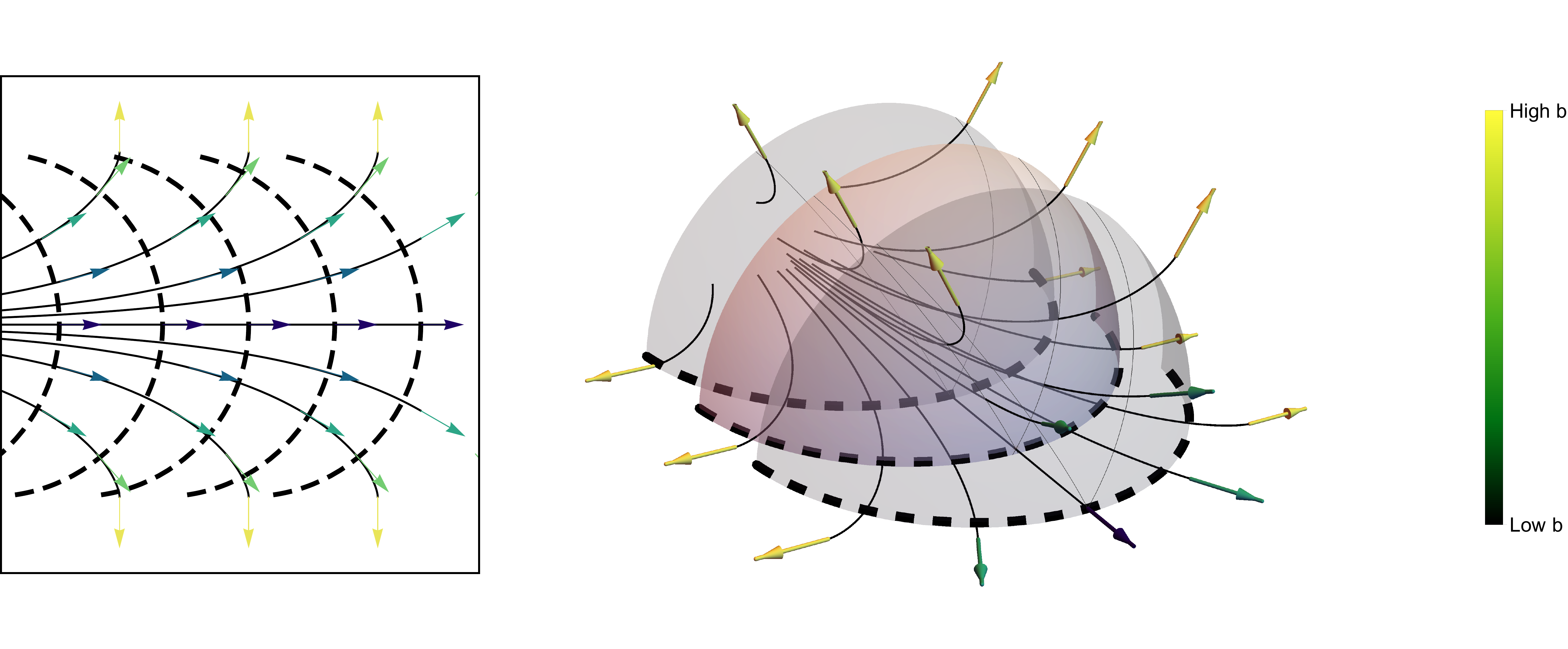}
    \caption{
    Example of a zero twist and zero biaxial splay phase. In the figures, the arrows representing the director are colored by the value of the bend $b$. (Left) The non-uniform bend and uniform splay $2d$ phase obtained by considering $\hatn$ orthogonal to the one-parameter family of circles of radius $\rho$ constant given by $\lambda\mapsto \mathbf{r}_{\lambda}(\theta)=(\lambda+\rho\cos\theta,\rho\sin\theta)$ (dashed  lines). The integral lines of $\hatn$ (full lines) have a bend equal to $b=\frac{1}{\rho}\tan\theta$ while the splay $s=2/\rho$ is constant. (Right) The $3d$ phase obtained by rotating the $2d$ phase on the left around $\bfr_{\lambda}$. The director $\hatn$ is orthogonal to a family of spheres with distinct centers but the same radii. Since $\hatn$ can be used as the unit normal of foliation of {a finite domain of} space by the spheres $\rho\mapsto\mathbf{R}_{\lambda}(\theta,\psi)=(\lambda+\rho\cos\theta,\rho\sin\theta\cos\psi,\rho\sin\theta\sin\psi)$, the twist vanishes $t=0$. In addition, as the leaves are totally umbilical surfaces, their shape operator $A_n=-\rmd\hatn$ are a multiple of the identity, see Eq. \eqref{Def::FamilyShapeOperators}. Therefore, the biaxial splay vanishes as well. 
    (The spheres have distinct colors to ease the visualization.)}
    \label{fig:3dHatNTDzeroSconstNon-UniformB}
\end{figure}

\begin{figure}[t]
    \centering
    \includegraphics[width=\linewidth]{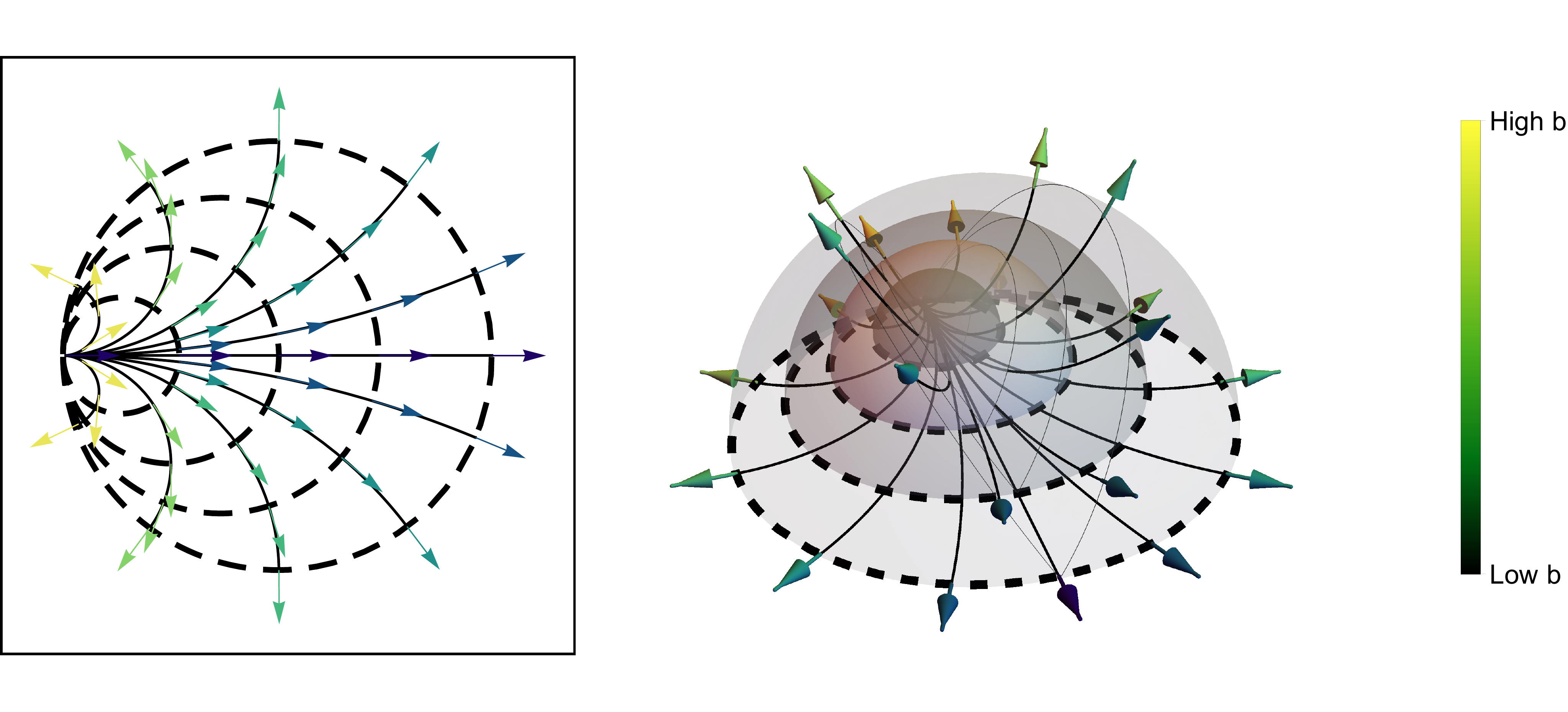}
    \caption{Example of a zero twist and zero biaxial splay phase. In the figures, the arrows representing the director are colored by the value of the bend $b$. (Left) The non-uniform bend and non-uniform splay $2d$ phase obtained by considering $\hatn$ orthogonal to the one-parameter family of circles $\lambda\mapsto \mathbf{r}_{\lambda}(\theta)=(\lambda-\lambda\cos\theta,\lambda\sin\theta)$ (dashed  lines). The integral lines of $\hatn$ (full lines) are circles or a horizontal line and, therefore, $\hatn\cdot\nabla b=0$. On the other hand, the splay varies only along the direction of $\hatn$: $\hatp\cdot\nabla s=0$. The derivatives $\hatp\cdot\nabla b$ and $\hatn\cdot\nabla s$ must comply with the $2d$ compatibility equation  $0=b^2+s^2+\hatn\cdot\nabla s+\hatp\cdot\nabla b$  \cite{NE18,NEcomment}. (Right) The $3d$ phase obtained by rotating the $2d$ phase on the left around $\bfr_{\lambda}$. The director $\hatn$ is orthogonal to a family of spheres with distinct centers and radii. Since $\hatn$ can be used as the unit normal of foliation by the spheres $\lambda\mapsto\mathbf{R}_{\lambda}(\theta,\psi)=(\lambda-\lambda\cos\theta,\lambda\sin\theta\cos\psi,\lambda\sin\theta\sin\psi)$, the twist vanishes $t=0$. As the leaves are totally umbilical surfaces, their shape operator $A_n=-\rmd\hatn$ is a multiple of the identity; see Eq. \eqref{Def::FamilyShapeOperators}. Therefore, the biaxial splay vanishes as well. (The spheres have distinct colors to ease the visualization.)}
    \label{fig:3dHatNTDzeroNon-UniformSB}
\end{figure}

If the biaxial splay vanishes, then the shape operator of the leaves orthogonal to the director is diagonal with two identical eigenvalues. Therefore, the leaves must be totally umbilical; they are either portions of spheres or planes \cite{DoCarmo76}, Prop. 4 of Chap. 3. It follows that the splay is constant along the leaves, i.e., $\hatp\cdot\nabla s = 0$ and $\hatq\cdot\nabla s = 0$. 

In Fig. \ref{fig:3dHatNTDzeroSconstNon-UniformB}, we illustrate a phase where the director $\hatn$ is orthogonal to the leaves of a foliation of a finite domain of space by spheres of constant radius centered along a line. In contrast,  Fig. \ref{fig:3dHatNTDzeroNon-UniformSB} illustrates a phase where the director $\hatn$ is orthogonal to the leaves of a foliation by spheres centered along a line {but with distinct radii. Finally, note that} concentric circles yield the hedgehog phase as illustrated in Fig. \ref{fig:HiliconicalndHedgehog}, Right.  

It turns out that the process of obtaining phases with zero twist and zero biaxial splay from the unit normals of a {foliation of an open set of $\mathbb{R}^3$ by pieces of spheres is generic}. Indeed, given a point $p$ in space, consider the integral line $c(u)$ of $\hatn$ passing through $p$. Then, every point of $c(u)$ is intersected by a unique sphere whose center and radius may be denoted by $\bfr(u)$ and $\rho(u)$. (If some leaves are planes, we should allow for $1/\rho$ to have zeros.) In Fig. \ref{fig:3dHatNTDzeroSconstNon-UniformB}, we have $\bfr(\lambda)=\lambda \hat{\mathbf{x}}$ and $\rho$ constant, while in Fig. \ref{fig:3dHatNTDzeroNon-UniformSB} we have $\bfr(\lambda)=\lambda \hat{\mathbf{x}}$ and $\rho(\lambda)=\lambda$.

Now, let us compute the deformation modes of phases with $\hatn$ orthogonal to a foliation by spheres. 
As concentric circles necessarily yield the well-known hedgehog or the trivial nematic phases, we may assume that the curve $\bfr$ describing the centers of the spheres does not degenerate to a point. Consider $\mathbf{r}(\lambda)$ parametrized by its arc-length, i.e., $\bfr'(\lambda)\cdot\bfr'(\lambda)=1$. Now, consider the Frenet frame of $\bfr$ \cite{struik}, $\{\hatT,\hatN,\hatB\}$, and define the frame
\[
 \hatn =\cos\theta\hatT+\sin\theta\cos\psi\hatN+\sin\theta\sin\psi\hatB,
\]
\[
 \hatp = -\sin\theta\hatT+\cos\theta\cos\psi\hatN+\cos\theta\sin\psi\hatB,  \quad \mbox{and} \quad
       \hatq = -\sin\psi \hatN+\cos\psi\hatB.
\]
Let us denote by $\kappa$ and $\tau$ the curvature and torsion of $\bfr$. 

If the director $\hatn$ is orthogonal to the spheres $\Sigma_{\lambda}=\mathbb{S}^2(\bfr(\lambda),\rho(\lambda))$ of radii $\rho(\lambda)$ and centered at the points of $\mathbf{r}(\lambda)$, then we can parametrize a neighborhood of space where $\hatn$ is defined as
\begin{equation}
    \mathbf{R}(\lambda,\theta,\psi) = \mathbf{r}(\lambda)+\rho(\lambda) \hatn(\lambda,\theta,\psi).
\end{equation}
We can write the vector fields in the frame $\{\hatn,\hatp,\hatq\}$ in terms of the parametric velocity vectors $\{\partial\bfR/\partial\lambda,\partial\bfR/\partial\theta,\partial\bfR/\partial\psi\}$  as
\begin{equation}
    \begin{array}{c}
         \hatn = \frac{1}{\rho'+\cos\theta}\frac{\partial\bfR}{\partial\lambda}+\frac{\sin\theta-\rho\kappa\cos\psi}{\rho'+\cos\theta}\frac{\partial\bfR}{\partial\theta}+\frac{\kappa\cos\theta\sin\psi-\tau\sin\theta}{\sin\theta(\rho'+\cos\theta)}\frac{\partial\bfR}{\partial\psi},  \\[6pt]
        \hatp = \frac{1}{\rho}\frac{\partial\bfR}{\partial\theta}, \quad \mbox{and} \quad \hatq = \frac{1}{\rho\sin\theta}\frac{\partial\bfR}{\partial\psi}. \\
    \end{array}
\end{equation}
Therefore, we can compute the directional derivatives $\hatn\cdot\nabla$, $\hatp\cdot\nabla$, and $\hatq\cdot\nabla$ in terms of the partial derivatives with respect to the coordinates $(\lambda,\theta,\psi)$:  recall that $\frac{\partial\bfR}{\partial\lambda}\cdot\nabla=\frac{\partial}{\partial\lambda}$, $\frac{\partial\bfR}{\partial\theta}\cdot\nabla=\frac{\partial}{\partial\theta}$, and $\frac{\partial\bfR}{\partial\psi}\cdot\nabla=\frac{\partial}{\partial\psi}$. Now, using Eqs. \eqref{eq::EvolEqsFrameNPQandPQN} and \eqref{eq::EvolEqsFrameQNP}, we can finally obtain that 
\begin{equation}
    b_{p} = -\dfrac{\sin\theta}{\rho(\cos\theta+\rho')},\, b_{q} = 0,\, \frac{s}{2}+\Delta_1 = \frac{1}{\rho},\, \frac{s}{2}-\Delta_1=\frac{1}{\rho},\, \frac{t}{2}-\Delta_2 = 0,\, \frac{t}{2}+\Delta_2 = 0. 
\end{equation}

Thus, this process provides us with phases of zero twist, zero biaxial splay, and splay $s=\frac{2}{\rho}$ uniform on each leaf as expected. Note that the bend displays non-trivial variation both within the leaves and between leaves as a function of $\rho(\lambda)$, yet retains the rotational symmetry in the plane normal to $\hatT$. 

\subsection{Vanishing bend phases}
\label{subsect::ZeroBend}

Next, we come to study directors with no bend, often called Beltrami fields. (Note that we do not assume  $\nabla\cdot\hatn=0$ as some authors do
in studying Beltrami fields \cite{CK20}.) We will show that this family of directors depends on three functions to be prescribed on an initial surface transversal to $\hatn$: see Eqs. \eqref{eq::SolutionSplayForBzero}, \eqref{eq::SolutionTwistForBzero}, and \eqref{eq::SolutionSadSplayForBzero} below. In addition, these three functions must comply with two differential relations as a consequence of Eqs. \eqref{eq::CompCondUsingABCR1223Bzero} and \eqref{eq::CompCondUsingABCR1323Bzero} below. (The solutions \eqref{eq::SolutionSplayForBzero}, \eqref{eq::SolutionTwistForBzero}, and \eqref{eq::SolutionSadSplayForBzero} are obtained from the compatibility equations \eqref{eq::CompCondUsingABCR1212Bzero}, \eqref{eq::CompCondUsingABCR1213Bzero}, \eqref{eq::CompCondUsingABCR1312Bzero}, and \eqref{eq::CompCondUsingABCR1313Bzero}, that only contain derivatives in the direction of $\hatn$.)

Setting $b_p=b_q=0$ in the compatibility equations \eqref{eq::CompCondUsingABCR1212}--\eqref{eq::CompCondUsingABCR1323} gives
\begin{eqnarray}
0 &=& -(\frac{s}{2}+\Delta_1)_{,n}-\frac{s^2}{4}+\frac{t^2}{4}-s\Delta_1-(\Delta)^2+2\alpha \Delta_2, \label{eq::CompCondUsingABCR1212Bzero} \\[3pt]
0 &=& -(-\frac{t}{2}+\Delta_2)_{,n}-s(-\frac{t}{2}+\Delta_2)-2\alpha \Delta_1, \label{eq::CompCondUsingABCR1213Bzero} \\[3pt]
0 &=& -(-\frac{t}{2}+\Delta_2)_{,p}+(\frac{s}{2}+\Delta_1)_{,q}-2\beta \Delta_1-2\gamma\Delta_2, \label{eq::CompCondUsingABCR1223Bzero} \\[3pt]
0 &=& -(\frac{t}{2}+\Delta_2)_{,n}-s(\frac{t}{2}+\Delta_2)-2\alpha \Delta_1, \label{eq::CompCondUsingABCR1312Bzero} \\[3pt]
0 &=& -(\frac{s}{2}-\Delta_1)_{,n}-\frac{s^2}{4}+\frac{t^2}{4}+s\Delta_1-(\Delta)^2-2\alpha \Delta_2, \label{eq::CompCondUsingABCR1313Bzero} \\[3pt]
0 &=& -(\frac{s}{2}-\Delta_1)_{,p}+(\frac{t}{2}+\Delta_2)_{,q}-2\beta \Delta_2+2\gamma\Delta_1. \label{eq::CompCondUsingABCR1323Bzero}
\end{eqnarray}
The coefficients $\alpha$, $\beta$, and $\gamma$ are
\begin{eqnarray}\label{eq::AlphaBeltramiFields}
    \alpha & = & \frac{\Delta_2\Delta_{1,n}-\Delta_1\Delta_{2,n}}{2\Delta^2},\label{eq::AlphaAsFunctionOfDeformationModesBzero}
\end{eqnarray}
\begin{eqnarray}
    \beta & = & \frac{\Delta_1t_{,p}-\Delta_2s_{,p}+\Delta_1s_{,q}+\Delta_2t_{,q}}{4\Delta^2}-\frac{\Delta_1\Delta_{2,p}-\Delta_2\Delta_{1,p}}{2\Delta^2}+\frac{(\Delta^2)_{,q}}{4\Delta^2},\label{eq::BetaAsFunctionOfDeformationModesBzero}
\end{eqnarray}
and
\begin{eqnarray}
    \gamma & = & \frac{\Delta_2t_{,p}+\Delta_1s_{,p}-\Delta_1t_{,q}+\Delta_2s_{,q}}{4\Delta^2}+\frac{\Delta_2\Delta_{1,q}-\Delta_1\Delta_{2,q}}{2\Delta^2}-\frac{(\Delta^2)_{,p}}{4\Delta^2}.\label{eq::GammaAsFunctionOfDeformationModesBzero}
\end{eqnarray}

Our main goal here is to show how the description of Beltrami fields depends on the information prescribed on an initial surface. In the present case, the triplet of deformation modes $\{t,s,\sigma\}$ can be interpreted more transparently and naturally and thus
 will be used instead of the triplet  $\{t,s,\Delta\}$.
Indeed, as discussed in Subsect. \ref{subsectZeroTwist}, the splay plays the role of the mean curvature, the saddle-splay plays the role of the Gaussian curvature, and the twist measures the deviation of the director from being orthogonal to the leaves of a foliation.

Summing Eq. \eqref{eq::CompCondUsingABCR1212Bzero} and Eq. \eqref{eq::CompCondUsingABCR1313Bzero} and then using Eq. \eqref{Eq::SigmaAsFunctionOfSTD}, we obtain
\begin{equation}\label{delnSBeltramiFields}
    s_{,n} + s^2 = \sigma.
\end{equation}
On the other hand, subtracting Eq. \eqref{eq::CompCondUsingABCR1312Bzero} from Eq. \eqref{eq::CompCondUsingABCR1213Bzero} gives
\begin{equation}\label{delnTBeltramiFields}
    t_{,n} + s t =0.
\end{equation}

To find the evolution equation for $\sigma$, we may use the evolution equation for $\Delta^2$. First, subtract Eq. \eqref{eq::CompCondUsingABCR1212Bzero} from Eq. \eqref{eq::CompCondUsingABCR1313Bzero}:
\begin{equation}\label{delnD1BeltramiFields}
    \Delta_{1,n}+s\Delta_1-2\alpha\Delta_2 = 0.
\end{equation}
Now, summing Eq. \eqref{eq::CompCondUsingABCR1312Bzero} and Eq. \eqref{eq::CompCondUsingABCR1213Bzero}:
\begin{equation}\label{delnD2BeltramiFields}
    \Delta_{2,n}+s\Delta_2+2\alpha \Delta_1 = 0.
\end{equation}
Finally, summing  Eq. \eqref{delnD1BeltramiFields} multiplied by $\Delta_1$ and Eq. \eqref{delnD2BeltramiFields} multiplied by $\Delta_2$ gives
\begin{equation}\label{delnDBeltramiFields}
    (\Delta^2)_{,n}+2s\Delta^2 = 0.
\end{equation}
Thus, using Eq. \eqref{Eq::SigmaAsFunctionOfSTD} together with Eqs. \eqref{delnSBeltramiFields} and \eqref{delnTBeltramiFields} allow us to obtain
\begin{equation}\label{delnSigmaBeltramiFields}
    \sigma_{,n}+s\sigma = 0.
\end{equation}

\begin{remark}
By choosing $\hatp$ and $\hatq$ such that $\alpha=0$ (see Remark \ref{remark::AlphaZero}), we obtain the evolution equation for $\Delta_i$ as $\Delta_{i,n}+s\Delta_i=0$, $i=1,2$. In addition, if $\alpha=0$, then Eq. \eqref{eq::AlphaBeltramiFields} implies that ratio between $\Delta_1$ and $\Delta_2$ is constant. In the case of Beltrami fields, we get $\alpha=0$ if we choose $\hatp$ and $\hatq$ as the eigenvectors of the biaxial splay: $\Delta_2=0\Rightarrow \alpha=0$ by Eq. \eqref{eq::AlphaAsFunctionOfDeformationModesBzero}.
\end{remark}

Let us integrate Eqs. \eqref{delnSBeltramiFields}, \eqref{delnTBeltramiFields}, and \eqref{delnSigmaBeltramiFields}. First, introduce a coordinate system $(u,v,w)$ such that $\hatn=\frac{\partial}{\partial u}$. For example, if we consider a surface $(v,w)\mapsto\mathbf{r}(v,w)$ transversal to $\hatn$, then we may define $(u,v,w)\mapsto\mathbf{R}(u,v,w)=\mathbf{r}(v,w)+u\,\hatn(v,w)$, where $\hatn(v,w)$ is the restriction of $\hatn$ on the points of $\mathbf{r}$. Then, $f_{,n}=\hatn\cdot\nabla f=\partial f/\partial u=f_{u}$. 

Taking the derivative of Eq. \eqref{delnSBeltramiFields} and using Eq. \eqref{delnSigmaBeltramiFields} imply $s_{uu}+2ss_u=\sigma_u=-s\sigma$. Using Eq. \eqref{delnSigmaBeltramiFields} again to eliminate $\sigma$ finally gives
\begin{equation}
s_{uu}+3ss_{u}+s^3=0.   
\end{equation}
Introducing $Y(s)=\mathrm{d}s/\mathrm{d}u$, the equation above is mapped onto $Y'(s)=-3s-s^3/Y(s)$, which is an equation of Chini type \cite{Kamke1948}, Eq. I$\cdot55$, p. 303. The general solution for $s$ is
\begin{equation}
    s(u,v,w)=\frac{2A(u+B)}{A\,u^2+2AB\,u+AB^2+3}=\frac{2A(u+B)}{A(u+B)^2+3},
\end{equation}
where $A=A(v,w)$ and $B=B(v,w)$. For $t$ and $\sigma$, the solutions are
\begin{equation}
    t(u,v,w)=t_0(v,w)\exp(-\int_0^u s(x)\mathrm{d}x)=\frac{C}{A(u+B)^2+3}
\end{equation}
and
\begin{equation}
    \sigma(u,v,w)=\sigma_0(v,w)\exp(-\int_0^u s(x)\mathrm{d}x)=\frac{D}{A(u+B)^2+3},
\end{equation}
where $C=C(v,w)$ and $D=D(v,w)$.

The functions $A,B,C$, and $D$ are not entirely arbitrary. Indeed, we have a few relations between the functions $A,B,C,D$ and $t,s$, and $\sigma$ at $u=0$, i.e., along the transversal surface $\mathbf{r}$. Namely, $s_0=2AB/(AB^2+3)$, $t_0=C/(AB^2+3)$, and $\sigma_0=D/(AB^2+3)$, where $s_0=s_0(v,w)=s(0,v,w)$, $t_0=t_0(v,w)=t(0,v,w)$, and $\sigma_0=\sigma_0(v,w)=\sigma_0(0,v,w)$. An additional equation comes from the fact that $s_0$ and $\sigma_0$ must be connected by $s_u+s^2=-\sigma$:
\begin{equation}
    s_u+s^2=\frac{2A}{A(B+u)^2+3},\sigma = \frac{D}{A(u+B)^2+3}\stackrel{u=0}{\Rightarrow} D=-2A.
\end{equation}
Now, dividing $s_0$ by $\sigma_0$ gives 
$$\frac{s_0}{\sigma_0}=\frac{2AB}{D}=-B.$$
In addition, using the expression for $\sigma_0$, we have 
\[
-2A=AB^2\sigma_0+3\sigma_0=As_0^2/\sigma_0+3\sigma_0,
\]
which gives
\[
A=-\frac{3\sigma_0}{2+\frac{s_0^2}{\sigma_0}}=-\frac{3\sigma_0^2}{2\sigma_0+s_0^2}\Rightarrow D=\frac{6\sigma_0^2}{2\sigma_0+s_0^2}.
\]
Finally, the remaining coefficient is
\[
C=t_0(AB^2+3)=t_0\left(-\frac{3\sigma_0^2}{2\sigma_0+s_0^2}\frac{s_0^2}{\sigma_0^2}+3\right)=\frac{6\,t_0\sigma_0}{2\,\sigma_0+s_0^2}.
\]

In short, the general solutions for the splay, twist, and saddle splay of a Beltrami director field are
\begin{equation}\label{eq::SolutionSplayForBzero}
    s = \frac{2\frac{-3\sigma_0^2}{2\sigma_0+s_0^2}(u-\frac{s_0}{\sigma_0})}{\frac{3[2\sigma_0+s_0^2-(s_0-u\sigma_0)^2]}{2\sigma+s_0^2}}=\frac{2\sigma_0(s_0-\sigma_0u)}{2\sigma_0+s_0^2-(s_0-\sigma_0u)^2}=\frac{s_0-\sigma_0u}{1+s_0u-\sigma_0\frac{u^2}{2}},
\end{equation}    
\begin{equation}\label{eq::SolutionTwistForBzero}    
    \,t=\frac{2t_0\sigma_0}{2\sigma_0+s_0^2-(s_0-\sigma_0u)^2}=\frac{t_0}{1+s_0u-\sigma_0\frac{u^2}{2}},
    \end{equation}
and    
\begin{equation}\label{eq::SolutionSadSplayForBzero}
    \sigma=\frac{2\sigma_0^2}{2\sigma_0+s_0^2-(s_0-\sigma_0u)^2}=\frac{\sigma_0}{1+s_0u-\sigma_0\frac{u^2}{2}}.
\end{equation}
As discussed after Eqs. \eqref{delnD1BeltramiFields} and \eqref{delnD2BeltramiFields}, we may choose $\hatp$ and $\hatq$ such that $\Delta_2=0$. Under this Gauge choice, there are only three remaining deformation modes to investigate, either $\{s,t,\sigma\}$ or $\{s,t,\Delta\}$. Remember that splay, twist, biaxial splay, and saddle-splay are connected by Eq. \eqref{Eq::SigmaAsFunctionOfSTD}.

The solutions \eqref{eq::SolutionSplayForBzero}, \eqref{eq::SolutionTwistForBzero}, and \eqref{eq::SolutionSadSplayForBzero} depend on three functions to be prescribed on an initial surface transversal to $\hatn$. However, these functions are not entirely arbitrary, as they must comply with Eqs. \eqref{eq::CompCondUsingABCR1223Bzero} and \eqref{eq::CompCondUsingABCR1323Bzero}.

Finally, note that  splay, twist, and saddle splay may be singular at those points $\mathbf{R}(u,v,w)$ where the denominator in Eqs. (\ref{eq::SolutionSplayForBzero}), (\ref{eq::SolutionTwistForBzero}), and (\ref{eq::SolutionSadSplayForBzero}) vanishes: $d(u)=1+s_0u-\sigma_0u^2/2=0$. If $\sigma_0=0$, this happens when $u=1/s_0$. On the other hand, if $\sigma_0>0$, then $s_0^2+2\sigma_0>0$ and $d(u)=0$ has two real distinct solutions, one positive and one negative. Indeed, $\sqrt{s_0^2+2\sigma_0}>\vert s_0\vert$ and then
\begin{equation}
\left\{
    \begin{array}{cc}
         s_0\geq0 \Rightarrow & \frac{s_0+\sqrt{s_0^2+2\sigma_0}}{\sigma_0}>\frac{s_0}{\sigma_0}+\frac{s_0}{\sigma_0}\geq0 \mbox{ and } \frac{s_0-\sqrt{s_0^2+2\sigma_0}}{\sigma_0}<\frac{s_0}{\sigma_0}-\frac{s_0}{\sigma_0}<0\\[4pt]
         s_0<0 \Rightarrow & \frac{s_0+\sqrt{s_0^2+2\sigma_0}}{\sigma_0}>\frac{s_0}{\sigma_0}-\frac{s_0}{\sigma_0}>0 \mbox{ and }  \frac{s_0-\sqrt{s_0^2+2\sigma_0}}{\sigma_0}<\frac{s_0}{\sigma_0}+\frac{s_0}{\sigma_0}<0\\ 
    \end{array}
    \right..
\end{equation}
Thus, for $\sigma_0<0$, $d(u)$ may have zero, one, or two real solutions as long as $s_0^2+2\sigma_0<0$, $=0$, or $>0$, respectively. When $s_0^2+2\sigma_0=0$, the unique solution for $d(u)=0$ is $u=s_0/\sigma_0$. Contrarily to the case $\sigma_0>0$, when $s_0^2+2\sigma_0>0$ and $\sigma_0<0$, the two solutions have the same sign.

As an example, consider the director field $\hat{\mathbf{n}}(x,y,z)=(\cos f(z),\sin f(z), 0)$, where $x,y,z$ are Cartesian coordinates and $f$ is a smooth function. Here, $\nabla\times\hat{\mathbf{n}}=(-f'\cos f,-f'\sin f,0)=-f'\hat{\mathbf{n}}$ and then $\mathbf{b}=\hat{\mathbf{n}}\times\nabla\times\hat{\mathbf{n}}=0$. Therefore,  $\hat{\mathbf{n}}$ is a Beltrami director field. The twist is $t=\hat{\mathbf{n}}\cdot\nabla\times\hat{\mathbf{n}}=-f'$, the splay is $s=\nabla\cdot\hat{\mathbf{n}}=0$, and finally the saddle splay is $\sigma=0$. {Let $\Sigma$ be a surface transversal to $\hatn$ as, for example, the plane $y=0$. Take as initial conditions on $\Sigma$ the values} $s_0(v,w)=0$, $\sigma_0(v,w)=0$, and {$t_0(v,w)=-f'(w)$}. As expected, applying Eqs. (\ref{eq::SolutionSplayForBzero}), (\ref{eq::SolutionTwistForBzero}), and (\ref{eq::SolutionSadSplayForBzero}) to these initial conditions gives {$t=t_0=-f'(w)$}, $s=0$, and $\sigma=0$.

\subsection{Stacked planar phases}

We conclude by examining the special case where the director constitutes stacked planar phases, yet with a possible twist. In this case, space is foliated by planes in which $\hatn$ and $\mathbf{b}$ span the local tangent planes. Setting $\hatp\parallel \mathbf{b}$ we obtain that $\hatq=\mbox{const.}$, and consequently $J^q=0$. We may thus deduce that   
\begin{equation}
\Delta_1=\frac{s}{2},\quad
\Delta_2=-\frac{t}{2},\quad
b_q=0,\quad\alpha=0,\quad
\beta=0,\quad
\gamma=0.
\end{equation}
In this case, the compatibility conditions assume a straightforward form:
\begin{equation}
    s^2+b_p^2+s_{,n}+b_{p,p}=0
    \label{eq:planarComp}
\end{equation}
 and 
 \begin{equation}
     b_{p,q}=s t+t_{,n} \,,\qquad
     s_{,q}=-b_p t-t_{,p} \, .
     \label{eq:planarNormalAdvect}
 \end{equation}
Equation \eqref{eq:planarComp} is the known two-dimensional compatibility condition within each of the leaves \cite{NE18,NEcomment}, while equations \eqref{eq:planarNormalAdvect} describe how the bend and splay evolve between leaves as a function of the twist and its tangential derivatives. We note that differentiating equation \eqref{eq:planarComp} along $\hatq$ and substituting equation  \eqref{eq:planarNormalAdvect} (along with the relations $(\hatq\cdot \nabla) \hatp=t\hatn$ and $(\hatq\cdot \nabla) \hatn=-t\hatp$) results in a trivial relation. Thus the satisfaction of equation \eqref{eq:planarComp} in one of the tangent planes is propagated along $\hatq$ by equations \eqref{eq:planarNormalAdvect} to all of space. We conclude that it suffices to prescribe the values the twist assumes in a three-dimensional domain and compatible splay and bend functions along a single plane to determine the domain's texture uniquely. The cholesteric phase described at the end of the previous subsection on Beltrami fields is a particular example in which the bend and splay vanish identically, and the twist is uniform across each planar leaf.     

\section{Concluding remarks}
\label{sec::Conclusion}

In this work, we derived through vector calculus the compatibility conditions for three-dimensional director fields in {Euclidean space}. The results presented here agree with the results obtained for uniform distortion fields \cite{V19} and also coincide with the results obtained through the method of moving frames \cite{dSE21,PA21}. {Our strategy consisted in seeing Eqs. \eqref{eq::deliNj}, \eqref{eq::deliPj}, and \eqref{eq::deliQj} as a system of PDEs defining an orthonormal triad containing the director. This system is naturally subjected to integrability conditions, which gives us compatibility equations. %\luiz{Conversely, in the appropriate context, e.g., Exterior Differential Systems, the compatibility equations provide sufficient conditions for a triad containing the director with prescribed deformation modes to exist. (We briefly explained this context in Appendix \ref{AppendixCompEqsAsSufficient}.)}

We have shown that every three-dimensional director field is fully characterized by five fields, namely $s,t,b,\Delta$, and $\phi$ (the splay, twist, bend magnitude, biaxial splay magnitude, and the relative angle between the bend vector and the principal directions of the biaxial splay, respectively).  The values these fields can assume in space are related to each other through six differential relations termed the compatibility conditions. %Any set of local fields that satisfies these equations corresponds to a realizable triad and a liquid crystalline texture with the matching values of $s,t,b,\Delta$, and $\phi$. 

{The existence of compatibility equations naturally leads to finding solutions to the set of compatibility conditions.} The most comprehensive way of finding solutions to equations \eqref{eq::CompCondUsingABCR1212}-\eqref{eq::CompCondUsingABCR2323} is to consider them as partial differential equations for the fields $s,t,b,\Delta$, and $\phi$, and seek their solvability conditions in the most general sense, and possibly classify their solutions. %\luiz{Alternatively, we can employ differential forms and see the compatibility equations as an Exterior Differential System. This approach provides a way of interpreting in which sense Eqs. \eqref{eq::CompCondUsingABCR1212}-\eqref{eq::CompCondUsingABCR2323} could be seen as sufficient conditions. We do not know of any  attempt to attack these formidable tasks directly.}
{Another approach considers} the compatibility conditions as algebraic equations for the sought triad. This approach yields the Cartesian components of the triad as functions of the intrinsic deformation modes and the Cartesian components of their gradients. By construction, these will satisfy the compatibility conditions. However, calculated in this fashion, there is no assurance that the resulting triad matches the sought deformation modes, e.g., that the splay of the found director indeed coincides with the prescribed value of $s$. Requiring that the resulting triad be orthonormal and that its deformation modes' values coincide with the prescribed ones yields another set of compatibility conditions. 
This idea of interpreting the compatibility conditions as algebraic equations for the components of the sought vector fields was first implemented in $2d$, see Section 3 of \cite{NE18}, and later in $3d$, see Section 4 of \cite{PA21}. However, in neither case were the most general compatibility conditions obtained.

Given the generality of the compatibility equations, it may be naive to expect to solve them in general without severe simplifying assumptions. In practice, the compatibility conditions should be {seen as complementing the Euler-Lagrange equations} associated with an energy functional written in terms of the deformation modes. The system composed of Euler-Lagrange equations and compatibility conditions may be amenable to further analytical progress or  numerical solutions. The path taken here does not incorporate the Euler-Lagrange equations but rather follows a less general yet more applicable route. {In this work, we interpreted and exploited the compatibility equations as necessary conditions for the existence of an orthonormal triad. We imposed constraints that some of the deformation modes were constant or vanishing and then solved reduced compatibility conditions for the  remaining modes. This allowed us to simplify the compatibility conditions  and obtain the corresponding compatible phases. Table 1 summarizes these results.}

\begin{table}[t]
\centering
\large
\begin{tabular}{|p{2.0cm}|p{5.0cm}|}
\hline
\multicolumn{2}{|l|}{Phases of uniform distortion modes} \\ \hline
\hline
\multicolumn{1}{|c|}{$s=0,t=\pm2\Delta,b_p=\pm b_q$}  & $\hatn$ is tangent to a foliation of space by parallel helices \\ \hline
\hline
\multicolumn{2}{|l|}{Phases with a single non-uniform distortion mode} \\ \hline\hline
\multicolumn{1}{|c|}{$t,b,\Delta=0,\nabla s\not=0$}  & $\hatn$ normal to a foliation by concentric spheres (hedgehog) \\ \hline
\multicolumn{1}{|c|}{$t,b_q,\Delta_2=0,s=2\Delta_1$,$\nabla b\not=0$}  & $\hatn$ normal to a foliation {by  cylinders with parallel axes} \\ \hline
\multicolumn{1}{|c|}{$t=-2\Delta_2$,$s=2\Delta_1$,$b_q=0$,$\nabla b\not=0$}  & Special biaxial cholesteric \\ \hline \hline
\multicolumn{2}{|l|}{Phases of vanishing twist ($\hatn$ is orthogonal to a foliation}\\
\multicolumn{2}{|c|}{of a domain by surfaces)}\\ 
\hline\hline
\multicolumn{1}{|c|}{$t=0,b=0$}  & \multicolumn{1}{|l|}{leaves are parallel surfaces} \\ \hline
\multicolumn{1}{|c|}{$t=0,\Delta=0$}  & \multicolumn{1}{|l|}{leaves are spheres$/$planes} \\ \hline
\multicolumn{1}{|c|}{$t=0,s=0$}  & \multicolumn{1}{|l|}{leaves are minimal surfaces} \\ \hline \hline
\multicolumn{2}{|l|}{Phases of vanishing bend (Beltrami fields)} \\ \hline\hline
\multicolumn{1}{|c|}{$b=0$}  & $\hatn$ tangent to a foliation by straight lines \\ \hline
\end{tabular}
\vspace{5pt}
\caption{Families of director fields with
restricted deformation modes.}
\label{tab:adicaoZ4}
\end{table}
 
In a few of the cases presented here, the compatibility conditions reduce to a boundary value problem  allowing to identify the relevant degrees of freedom that fully determine the phase in space. These dimensionally reduced descriptions of the phases allow us to understand ``how many" distinct fields could be constructed with the desired property. More importantly, they also advance our understanding in approaching the relevant inverse design problems, where one seeks the local fields to prescribe in order to  obtain a specific desired texture \cite{GAE19}. 

The compatibility conditions not only identify which local deformation modes could not exist in $\mathbb{R}^3$ but can also  guide the construction of optimal compromises that approximate well these unattainable phases. 
The heliconical phase that arises in lieu of a phase of a constant bend and vanishing splay, twist, and saddle splay constitutes a uniform compromise associated with extensive energy. For small enough domains, one may expect non-uniform phases that are associated with super-extensive energy scaling \cite{ME21} to provide a better approximation for the unattainable phase. In this case, an optimal compromise could be constructed by coinciding with the desired incompatible deformation modes along a surface (or a curve) and using the compatibility conditions to  propagate the texture away from the surface (curve) along which the deformation modes were prescribed.

\begin{appendices}

\section {Proof of Theorem \ref{thrRigNonUnifDelta}}
\label{appendixProofThm1}

In what follows, we make use of the shorthand notations  
 \begin{equation}\label{def::DbDotb}
 \langle \mathcal{D}\mathbf{b},\mathbf{b}\rangle=(b_{p}^2-b_{q}^2)\Delta_1+2b_{p}b_{q}\Delta_2=b^2\Delta \cos(2\phi)    
\end{equation}
and 
 \begin{equation}\label{def::JDbDotb}
 \langle \mathcal{J}\mathcal{D}\mathbf{b},\mathbf{b}\rangle=(b_{q}^2-b_{p}^2)\Delta_2+2b_{p}b_{q}\Delta_1=b^2\Delta \sin(2\phi).    
\end{equation}
Here, $\phi$ is the angle formed by the bend vector and the principal direction of the biaxial splay, while $\mathcal{D}$ and $\mathcal{J}$ denote the biaxial splay and the counterclockwise $\frac{\pi}{2}$-rotation acting as linear operators on the plane normal to the director field, respectively. 

If $b_p,b_q,t$, and $s$ are constant and  $\Delta_2=0$, so $\Delta_1=\Delta$, then  Eqs. \eqref{eq::CompCondUsingABCR1212}--\eqref{eq::CompCondUsingABCR1323} become
\begin{eqnarray}
0 &=& -\Delta_{,n}-b_p^2-\frac{s^2}{4}+\frac{t^2}{4}-s\Delta-\Delta^2+\beta b_q, \label{eq::R1212BTSconst} \\
0 &=& -b_pb_q+\frac{st}{2}-2\alpha \Delta+\gamma b_q, \label{eq::R1213BTSconst}\\
0 &=& \Delta_{,q}+tb_p-2\beta \Delta, \label{eq::R1223BTSconst}\\
0 &=& -b_pb_q-\frac{st}{2}-2\alpha \Delta-\beta b_p, \label{eq::R1312BTSconst}\\
0 &=& \Delta_{,n}-b_q^2-\frac{s^2}{4}+\frac{t^2}{4}+s\Delta-\Delta^2-\gamma b_p,\label{eq::R1313BTSconst}\\
0 &=& \Delta_{,p}+tb_q+2\gamma \Delta. \label{eq::R1323BTSconst}
\end{eqnarray}

The coefficients $\alpha$, $\beta $, and $\gamma $ are given by
% \begin{equation}
%     \alpha = -\frac{tb^2}{8\Delta^2}-\frac{b_q\Delta_{,p}+b_p\Delta_{,q}}{8\Delta^2}-\frac{b_pb_q}{2\Delta},
% \end{equation}
% \begin{equation}
%     \beta  = \frac{(\Delta^2)_{,q}}{4\Delta^2}+t\,\frac{b_p\Delta}{2\Delta^2}=\frac{\Delta_{,q}}{2\Delta}+t\,\frac{b_p}{2\Delta},
% \end{equation}
\begin{equation}
    \alpha = -\frac{tb^2}{8\Delta^2}-\frac{b_q\Delta_{,p}+b_p\Delta_{,q}}{8\Delta^2}-\frac{b_pb_q}{2\Delta},
\,   \beta  = \frac{(\Delta^2)_{,q}}{4\Delta^2}+t\,\frac{b_p\Delta}{2\Delta^2}=\frac{\Delta_{,q}}{2\Delta}+t\,\frac{b_p}{2\Delta},
\end{equation}
and
\begin{equation}
    \gamma  = -\frac{(\Delta^2)_{,p}}{4\Delta^2}-t\,\frac{b_q\Delta}{2\Delta^2}=-\frac{\Delta_{,p}}{2\Delta}-t\,\frac{b_q}{2\Delta}.
\end{equation}

Now, using the above expressions for $\alpha$, $\beta$, and $\gamma$ and summing Eqs. \eqref{eq::R1212BTSconst} and \eqref{eq::R1313BTSconst} give
\begin{equation}
    0=\frac{b_p\Delta_{,p}+b_q\Delta_{,q}}{2\Delta}+\frac{-2\Delta b^2+2b_pb_qt-4\Delta^3-\Delta(s^2-t^2)}{2\Delta}.\label{eqp1}
\end{equation}
Subtracting Eq. \eqref{eq::R1312BTSconst} from Eq. \eqref{eq::R1213BTSconst}  gives
\begin{equation}
     0=\frac{-b_q\Delta_{,p}+b_p\Delta_{,q}}{2\Delta}+\frac{(b_p^2-b_q^2)t+2\Delta st}{2\Delta}.\label{eqp2}
\end{equation}
If, in addition, $b_p=b_q=0$, then $4\Delta^2+s^2-t^2=0$, from which we conclude that $\Delta$ must be constant. From now on, assume that $b^2\not=0$. Therefore, we get
\begin{equation}
    b^2\Delta_{,p} = 4 b_p\Delta^3+[ 2b_pb^2+(s^2-t^2)b_p+2stb_q]\,\Delta -b_q b^2t
\end{equation}
and
\begin{equation}
    b^2\Delta_{,q} = 4 b_q\Delta^3+[2b_qb^2+(s^2-t^2)b_q-2stb_p]\,\Delta -b_p b^2t.
\end{equation}

We shall now exploit Eq. \eqref{eq::CompCondUsingABCR2323}. First, we have
\begin{eqnarray}
    \beta_{,q} &=& \frac{\Delta\Delta_{,qq}-(\Delta_{,q})^2-tb_p\Delta_{,q}}{2\Delta^2}\nonumber\\
    & = & \frac{12 b_q\Delta^2+  2b_qb^2+(s^2-t^2)b_q-2stb_p}{2b^2\Delta}\Delta_{,q}-\frac{\Delta_{,q}+tb_p}{2\Delta^2}\Delta_{,q}\nonumber\\
    & = & \frac{4b_q \Delta}{b^4}\Big(4b_q\Delta^3+\Delta[2b_qb^2+(s^2-t^2)b_q-2stb_p]-b_pb^2t\Big)
\end{eqnarray}
and
\begin{eqnarray}
    -\gamma_{,p} &=& \frac{\Delta\Delta_{,pp}-(\Delta_{,p})^2-tb_q\Delta_{,p}}{2\Delta^2}\nonumber\\
    & = & \frac{12 b_p\Delta^2+ 2b_pb^2+(s^2-t^2)b_p+2stb_q}{2b^2\Delta}\Delta_{,p}-\frac{\Delta_{,p}+tb_q}{2\Delta^2}\Delta_{,p}\nonumber\\
    & = & \frac{4b_p \Delta}{b^4}\Big(4b_p\Delta^3+\Delta[2b_pb^2+(s^2-t^2)b_p+2stb_q]-b_qb^2t\Big).
\end{eqnarray}
Finally, from Eq. \eqref{eq::CompCondUsingABCR2323}, we obtain the following rational function in $\Delta$:
\[
\frac{12}{b^2\Delta^2}\Big\{\Delta^6+\frac{2(s^2-t^2)+5b^2}{12b^2}\Delta^4+ \frac{b^2(3t^2-4b^2-5s^2)- (s^2 + t^2)^2}{48}\Delta^2
\]
\begin{equation}
-\frac{tb_pb_q}{2}\Delta^3+\frac{tb_q [(s^2 - t^2) b_p + 2 s t b_q] - st^2b^2 + 
   3 tb_p b_q b^2}{24} \Delta-\frac{b^4t^2}{96}\Big\}=0.
\end{equation}
As the coefficient of the highest power never vanishes, it follows that $\Delta$ must be constant provided that $s,t,b_p$, and $b_q$ are constant.

\section {Proof of Theorem \ref{thrRigNonUnifBend}}
\label{appendixProofThm2}

If $\Delta_1,\Delta_2,t$, and $s$ are constant and $b_q=0$ (so, $b_p=b$), the compatibility equations \eqref{eq::CompCondUsingABCR1212}--\eqref{eq::CompCondUsingABCR1323} become
\begin{eqnarray}
0 &=& -b_{,p}-b^2-\frac{s^2}{4}+\frac{t^2}{4}-s\Delta_1-(\Delta)^2+2\alpha \Delta_2,\label{eq::R1212DeltaTSConst} \\
0 &=& -b_{,q}-s(-\frac{t}{2}+\Delta_2)-2\alpha\Delta_1,\label{eq::R1213DeltaTSConst} \\
0 &=& tb-2\beta \Delta_1-2\gamma \Delta_2,\label{eq::R1223DeltaTSConst} \\
0 &=& -s(\frac{t}{2}+\Delta_2)-2\alpha\Delta_1-\beta b,\label{eq::R1312DeltaTSConst}\\
0 &=& -\frac{s^2}{4}+\frac{t^2}{4}+s\Delta_1-(\Delta)^2-2\alpha\Delta_2-\gamma b,\label{eq::R1313DeltaTSConst}\\
0 &=& -2\beta \Delta_2+2\gamma \Delta_1. \label{eq::R1323DeltaTSConst}
\end{eqnarray}
If $\Delta_1^2+\Delta_2^2\not=0$, then the coefficients $\alpha$, $\beta $, and $\gamma $ are given by
\begin{equation}
    \alpha  = -\frac{\Delta_1b_{,q}-\Delta_2b_{,p}}{4\Delta^2}-\frac{tb^2}{8\Delta^2}-\frac{\langle \mathcal{J}\mathcal{D}\mathbf{b},\mathbf{b}\rangle}{4\Delta^2},\,\beta  = \frac{tb\Delta_1}{2\Delta^2},\, \mbox{ and } \,
   \gamma  = \frac{tb\Delta_2}{2\Delta^2}.
\end{equation}
% \begin{equation}
%     \beta  = \frac{tb\Delta_1}{2\Delta^2},\,
%    \gamma  = \frac{tb\Delta_2}{2\Delta^2},
% \end{equation}
% and
% \begin{equation}
%     \alpha  = -\frac{\Delta_1b_{,q}-\Delta_2b_{,p}}{4\Delta^2}-\frac{tb^2}{8\Delta^2}-\frac{\langle \mathcal{J}\mathcal{D}\mathbf{b},\mathbf{b}\rangle}{4\Delta^2}.
% \end{equation}
Here, we used the notation introduced in Eq. \eqref{def::JDbDotb}. Substituting $\alpha$, $\beta$, and $\gamma$ in Eqs. \eqref{eq::R1212DeltaTSConst} and  \eqref{eq::R1213DeltaTSConst} gives
\begin{equation}
    0 = \frac{\Delta_2^2-2\Delta^2}{2\Delta^2}b_{,p}-\frac{\Delta_1\Delta_2}{2\Delta^2}b_{,q}-b^2-\frac{s^2}{4}+\frac{t^2}{4}-s\Delta_1-\Delta^2-\frac{tb^2\Delta_2}{4\Delta^2}-\frac{\langle \mathcal{J}\mathcal{D}\mathbf{b},\mathbf{b}\rangle}{2\Delta^2}\Delta_2 \label{eq::AuxA}
\end{equation}
and
\begin{equation}\label{eq::AuxB}
0 =-\frac{\Delta_1\Delta_2}{2\Delta^2}b_{,p}+\frac{\Delta_1^2-2\Delta^2}{2\Delta^2}b_{,q}-s(-\frac{t}{2}+\Delta_2)+\frac{tb^2\Delta_1}{4\Delta^2}+\frac{\langle \mathcal{J}\mathcal{D}\mathbf{b},\mathbf{b}\rangle}{2\Delta^2}\Delta_1.
\end{equation}
Now, summing Eq. (\ref{eq::AuxA}) multiplied by $(\Delta_1^2-2\Delta^2)$ with Eq. (\ref{eq::AuxB}) multiplied by $\Delta_1\Delta_2$ gives
\begin{equation}\label{eq::bder2_TSD1D2Const}
     b_{,p} = \frac{2\Delta^2-\Delta_1^2}{\Delta^2}(\frac{t^2}{4}-\frac{s^2}{4}-\Delta^2)-b^2-\frac{tb^2}{2\Delta^2}\Delta_2-\frac{s\Delta_1\Delta_2}{\Delta^2}(\frac{t}{2}+\Delta_2)-\frac{s\Delta_1^3}{\Delta^2}.
\end{equation}
On the other hand, summing Eq. (\ref{eq::AuxA}) multiplied by $\Delta_1\Delta_2$ with Eq. (\ref{eq::AuxB}) multiplied by $(\Delta_2^2-2\Delta^2)$ gives
\begin{equation}\label{eq::bder3_TSD1D2Const}
     b_{,q} = s\frac{2\Delta^2-\Delta_2^2}{\Delta^2}(\frac{t}{2}-\Delta_2)+\frac{\Delta_1\Delta_2}{\Delta^2}(\frac{s^2}{4}-\frac{t^2}{4}+\Delta^2)+\frac{s\Delta_1^2\Delta_2}{\Delta^2}+\frac{tb^2}{2\Delta^2}\Delta_1.
\end{equation}

Let us now exploit the remaining compatibility equations, Eqs. \eqref{eq::CompCondUsingABCR2312}, \eqref{eq::CompCondUsingABCR2313}, and \eqref{eq::CompCondUsingABCR2323} . First, we can rewrite $\alpha $ as
\begin{eqnarray}
    \alpha  & = & \frac{1}{4\Delta^2}\left[2\Delta_2\left(\frac{t^2}{4}-\frac{s^2}{4}-\Delta^2\right)-b^2\left(\frac{t}{2}+\Delta_2\right)-st\Delta_1-\frac{tb^2}{2}+b^2\Delta_2\right]\nonumber\\
    & = & \frac{1}{4\Delta^2}\left[2\Delta_2\left(\frac{t^2}{4}-\frac{s^2}{4}-\Delta^2\right)-st\Delta_1-tb^2\right].
\end{eqnarray}
Then, Eq. \eqref{eq::CompCondUsingABCR2323} gives
\begin{eqnarray}
    0 & = & -\gamma_{,p}+\beta_{,q}-\beta^2-\gamma^2-t\,\alpha -\frac{s^2}{4}-\frac{t^2}{4}+(\Delta)^2\nonumber\\
    & = & \frac{t}{2\Delta^2}(\Delta_1b_{,q}-\Delta_2b_{,p})-\frac{t^2b^2}{4\Delta^2}-t\,\alpha -\frac{s^2}{4}-\frac{t^2}{4}+(\Delta)^2\nonumber\\
    & = & -\frac{t}{2\Delta^2}\left[2\Delta_2\left(\frac{t^2}{4}-\frac{s^2}{4}-\Delta^2\right)-b^2\left(\frac{t}{2}+\Delta_2\right)-st\Delta_1\right]-\frac{t^2b^2}{4\Delta^2}\nonumber\\
    &  &- \frac{t}{4\Delta^2}\left[2\Delta_2\left(\frac{t^2}{4}-\frac{s^2}{4}-\Delta^2\right)-st\Delta_1-tb^2\right]-\frac{s^2}{4}-\frac{t^2}{4}+\Delta^2\nonumber\\
    & = & \frac{tb^2}{2\Delta^2}\left(\frac{t}{2}+\Delta_2\right)- \frac{3t\Delta_2}{2\Delta^2}\left(\frac{t^2}{4}-\frac{s^2}{4}-\Delta^2\right)+\frac{3st^2\Delta_1}{4\Delta^2}-\frac{s^2}{4}-\frac{t^2}{4}+\Delta^2.\nonumber
\end{eqnarray}
Therefore, if neither $t=-2\Delta_2$ nor $t=0$, then $b$ must be constant.

It remains to analyze the case where $\Delta_1^2+\Delta_2^2=0$. Here, $\beta $ and $\gamma $ are given by
\begin{equation}
    \beta = -\frac{st}{2b}\mbox{ and }\gamma  = - \frac{s^2-t^2}{4b}.
\end{equation}
The Eqs. \eqref{eq::CompCondUsingABCR1212}, \eqref{eq::CompCondUsingABCR1213}, and \eqref{eq::CompCondUsingABCR1223} are \[
0 = -b_{,p}-b^2-\frac{s^2}{4}+\frac{t^2}{4},\,
0 = -b_{,q}+\frac{st}{2},\,
0 = tb.
\]
From the last equation, we conclude that either $b=0$ or $t=0$. Since we are assuming $t\not=0$, then $b=0$, which implies that $s=\pm t$ and $st=0$. We then deduce that $t$ must vanish, which is a contradiction. Consequently, no director field exists with $\Delta^2_1+\Delta_2^2=0$, $b=0$, $s=0$, but $t\not=0$. 

\section{Proof of Theorem \ref{thrRigNonUnifTwistOrSplay}}
\label{appendixProofThm3}

(a)  Let us assume that $b_p,b_q$, and $t$ are constant, and $\Delta_1^2+\Delta_2^2=0$. Then, Eqs. \eqref{eq::CompCondUsingABCR1212}--\eqref{eq::CompCondUsingABCR1323}  become 
\begin{eqnarray}
0 &=& -\frac{1}{2}s_{,n}-b_p^2-\frac{s^2}{4}+\frac{t^2}{4}+\beta b_q, \label{eq::R1212DeltaBTconst} \\
0 &=& -b_pb_q+\frac{st}{2}+\gamma b_q, \label{eq::R1213DeltaBTconst} \\
0 &=& \frac{1}{2}s_{,q}+tb_p, \label{eq::R1223DeltaBTconst} \\
0 &=& -b_pb_q-\frac{st}{2}-\beta b_p, \label{eq::R1312DeltaBTconst} \\
0 &=& -\frac{1}{2}s_{,n}-b_q^2-\frac{s^2}{4}+\frac{t^2}{4}-\gamma b_p, \label{eq::R1313DeltaBTconst} \\
0 &=& -\frac{1}{2}s_{,p}+tb_q. \label{eq::R1323DeltaBTconst}
\end{eqnarray}
If, in addition, we assume $b\not=0$, then we can write $\beta $ and $\gamma $ as
\begin{equation}
    \beta  = \frac{b_q}{2b^2} s_{,n} + \frac{(s^2 - t^2) b_q - 2 s t b_p}{4b^2},\,
    \gamma  = -\frac{b_p}{2b^2} s_{,n} - \frac{(s^2 - t^2) b_p + 2 s t b_q}{4b^2}.
\end{equation}
By summing Eqs. \eqref{eq::R1212DeltaBTconst} and \eqref{eq::R1313DeltaBTconst}, we obtain
\begin{equation}
    s_{,n} = -2 b^2 + \frac{t^2}{2} - \frac{s^2}{2},
\end{equation}
while from Eq. \eqref{eq::R1323DeltaBTconst} and \eqref{eq::R1223DeltaBTconst} we obtain
\begin{equation}
    s_{,p} = 2 t b_q\mbox{ and }s_{,q} = -2 t b_p.
\end{equation}

So far, we have not specified who $\hatp$ and $\hatq$ are. Choosing them to satisfy $\alpha=0$ (Remark \ref{remark::AlphaZero}), we finally obtain from Eq. \eqref{eq::CompCondUsingABCR2323} that
\begin{equation}
0=    -\frac{t^2 + b^2}{4b^2} s^2 + \frac{3 t^2 - 4 b^2}{4}.
\end{equation}
Since the coefficient of $s^2$ does not vanish, it follows that $s$ must be constant.

Finally, if $\Delta_1^2+\Delta_2^2=0$ and $b=0$, then we have
\begin{eqnarray}
0 &=& -\frac{s_{,n}}{2}-\frac{s^2}{4}+\frac{t^2}{4}, \label{eq::R1212TconstBDeltaZero} \\
0 &=& \frac{st}{2}, \label{eq::R1213TconstBDeltaZero} \\
0 &=& \frac{s_{,q}}{2}, \label{eq::R1223TconstBDeltaZero} \\ 
0 &=& -\frac{st}{2}, \label{eq::R1312TconstBDeltaZero} \\
0 &=& -\frac{s_{,n}}{2}-\frac{s^2}{4}+\frac{t^2}{4}, \label{eq::R1313TconstBDeltaZero} \\
0 &=& -\frac{s_{,p}}{2}. \label{eq::R1323TconstBDeltaZero}
\end{eqnarray}
Then, either $t=0$ or $s=0$. If $s=0$, then it also follows from Eq. \eqref{eq::R1212TconstBDeltaZero} that $t=0$ and, therefore, $\hatn$ is constant. On the other hand, if $t=0$ we deduce that $s$ only varies along the integral curves of $\hatn$ and satisfies $s_{,n}=-\frac{s^2}{2}$. This leads to a solution where $\hatn$ is either constant or a hedgehog, i.e., $\hatn$ corresponds to the unit normal field of a foliation of space by concentric spheres.

(b) Now, let us assume that $b_p,b_q$, and $s$ are constant, and $\Delta_1^2+\Delta_2^2=0$. Then, the compatibility equations \eqref{eq::CompCondUsingABCR1212}--\eqref{eq::CompCondUsingABCR1323} become 
 \begin{eqnarray}
 0 &=& -b_p^2-\frac{s^2}{4}+\frac{t^2}{4}+\beta b_q, \label{eq::R1212DeltaBSconst} \\
 0 &=& \frac{1}{2}t_{,n}-b_pb_q+\frac{st}{2}+\gamma b_q,\label{eq::R1213DeltaBSconst} \\
 0 &=& \frac{1}{2}t_{,p}+tb_p,\label{eq::R1223DeltaBSconst}\\
 0 &=& -\frac{1}{2}t_{,n}-b_pb_q-\frac{st}{2}-\beta b_p,\label{eq::R1312DeltaBSconst}\\
 0 &=& -b_q^2-\frac{s^2}{4}+\frac{t^2}{4}-\gamma b_p,\label{eq::R1313DeltaBSconst}\\
 0 &=& \frac{1}{2}t_{,q}+tb_q.\label{eq::R1323DeltaBSconst}
 \end{eqnarray}
If, in addition, we assume that $b\not=0$, then we can write $\beta $ and $\gamma $ as
\begin{equation}
    \beta  = -\frac{b_p}{2b^2} t_{,n} + \frac{(s^2 - t^2) b_q - 2 s t b_p}{4b^2},\,
    \gamma  = -\frac{b_q}{2b^2} t_{,n} - \frac{(s^2 - t^2) b_p + 2 s t b_q}{4b^2}.
\end{equation}
Finally, using $\beta$ and $\gamma$ above and summing Eqs. \eqref{eq::R1212DeltaBSconst} and \eqref{eq::R1313DeltaBSconst} give
\begin{equation}
    0 = b^2+\frac{t^2- s^2}{4},
\end{equation}
which implies that the twist must be constant. 

Finally, if $\Delta_1^2+\Delta_2^2=0$ and $b=0$, then Eq. \eqref{eq::R1212DeltaBSconst} becomes
 \[
 0= -\frac{s^2}{4}+\frac{t^2}{4}, 
 \]
which implies $t=\pm s$. Therefore, $t$ must be constant. This concludes the proof of Thm.  \ref{thrRigNonUnifTwistOrSplay}.

\section{Compatibility equations as sufficient conditions: director fields via Cartan's method of moving frames}
\label{AppendixCompEqsAsSufficient}

{The set of equations \eqref{eq::CompCondUsingABCR1212}--\eqref{eq::CompCondUsingABCR2323} provides necessary conditions for the existence of a solution of Eqs. \eqref{eq::deliNj}, \eqref{eq::deliPj}, and \eqref{eq::deliQj}. However, seeing them also as sufficient conditions seems nonsensical as they involve derivatives in the direction of the vectors we are trying to build. We can circumvent this difficulty by resorting to the formalism of Cartan's method of moving frames.}

%\luiz{Treating the compatibility equations as sufficient conditions for the existence of an orthonormal triad is possible, provided that we see them as an Exterior Differential System (EDS). Defining what it means to integrate such an EDS can be done with the help of Theorem \ref{thr::TenenblatLemma} below. While Theorem \ref{thr::TenenblatLemma} can be proved  using elementary tools, such as the Frobenius and Implicit Function Theorems, being able to solve Eqs. \eqref{eq::CompCondUsingABCR1212}--\eqref{eq::CompCondUsingABCR2323} for some large class of functions will likely require more powerful tools in the theory of EDSs, such as the Cartan-K\"ahler Theorem \cite{Bryant+91}.  }

%Treating the compatibility equations as sufficient conditions for the existence of an orthonormal triad is possible, provided that we use the formalism of differential forms. Now, we provide a context in which we can see Eqs. \eqref{eq::CompCondUsingABCR1212}--\eqref{eq::CompCondUsingABCR2323} as sufficient conditions for the existence of an orthonormal triad. 

{Given an orthonormal triad $\{\hatn_i\}_{i=1}^3$, we consider the set of dual 1-forms $\{\eta^i\}_{i=1}^3$: $\eta^i(\hatn_j)=\delta_j^i$. (See, e.g., Sect. 2 of Ref. \cite{dSE21} for background material.) The differential of $\hatn_i$ allows us to define new 1-forms $\eta_i^j$, the connection forms, by the relation $\rmd\hatn_i=\eta_i^k\hatn_k$ (sum on repeated indexes). Note $\hatn_i\cdot\hatn_j=\delta_{ij}\Rightarrow\eta_i^j=-\eta_j^i$. The 1-forms $\eta^j$ and $\eta_i^j$ are subject to the so-called Structure Equations}
\begin{equation}\label{StructureEqs}
 \rmd\eta^i = \eta^k\wedge\eta_k^i \quad \mbox{ and } \quad \rmd\eta_j^i = \eta_j^k\wedge\eta_k^i.
\end{equation}
{If $\hatn_1=\hatn$, $\hatn_2=\hatp$, and $\hatn_3=\hatq$, then we can write the connection forms $\eta_i^j$ as a linear combination of $\eta^i$ using the deformation modes (see Sect. 5 of Ref. \cite{dSE21}):}
\begin{equation}\label{eq::eta12UsingDefModes}
    \eta_1^2 = -b_p\eta^1 + \left(\frac{s}{2}+\Delta_1\right)\eta^2 + \left(-\frac{t}{2}+\Delta_2\right)\eta^3,
\end{equation}
\begin{equation}\label{eq::eta13eta23UsingDefModes}
 \eta_1^3 = -b_q\eta^1 + \left(\frac{t}{2}+\Delta_2\right)\eta^2 + \left(\frac{s}{2}-\Delta_1\right)\eta^3, \quad \mbox{and} \quad \eta_2^3 = \alpha\eta^1 + \beta\eta^2 + \gamma\eta^3.
\end{equation}
The compatibility equations \eqref{eq::CompCondUsingABCR1212}--\eqref{eq::CompCondUsingABCR2323} are then obtained by applying the structure equations to $\eta_i^j$ written as in Eqs. \eqref{eq::eta12UsingDefModes} and \eqref{eq::eta13eta23UsingDefModes}. (See Sect. 5 of Ref. \cite{dSE21}) Conversely. prescribing sets of 1-forms $\eta^i$ and $\eta_i^j=-\eta_j^i$ allows us to obtain a local existence theorem for an orthonormal triad $\hatn_i$.
\begin{theorem}\label{thr::TenenblatLemma}
 Let $\eta^i$ and $\eta_i^j=-\eta_j^i$ be differential 1-forms locally defined in $\mathbb{R}^3$ and assume they satisfy the structure equations \eqref{StructureEqs}. Then, there exists an orthonormal triad $\{\hatn_i\}_{i=1}^3$ such that $\eta^1,\eta^2$, and $\eta^3$ are the dual 1-forms and $\eta_i^j$ are the connection forms.
\end{theorem}
Proof of this theorem can be found as Lemma 2 of Ref. \cite{T71} in the case $k=0$. Tenenblat's proof is elementary and systematically uses differential forms and the Frobenius theorem. In particular, we may prescind the use of more powerful tools in the theory of exterior differential systems, such as the Cartan-K\"ahler Theorem \cite{Bryant+91}.

{In the context of director fields, we need to prescribe the 1-forms $\eta^i$, i.e., a basis of the cotangent space, together with the deformation modes $b_p$, $b_q$, $s$, $t$, $\Delta_1$, $\Delta_2$, $\alpha$, $\beta$, and $\gamma$. The prescription of the deformation modes allows us to define $\eta_i^j$ as in Eqs. \eqref{eq::eta12UsingDefModes} and \eqref{eq::eta13eta23UsingDefModes}, while prescribing the 1-forms $\eta^i$ allows us to make sense of the derivatives in the compatibility equations. More precisely, the directional  derivatives $f_{,n}=\hatn\cdot\nabla f$, $f_{,p}=\hatp\cdot\nabla f$, and $f_{,q}=\hatq\cdot\nabla f$ must be interpreted as the coordinates of the differential $\rmd f$ in the prescribed basis $\{\eta^1,\eta^2$, $\eta^3\}$. With this proviso, satisfying the structure equations \eqref{StructureEqs} requires the validity of Eqs. \eqref{eq::CompCondUsingABCR1212}--\eqref{eq::CompCondUsingABCR2323}.}

{Seeing the  compatibility equations as sufficient conditions for the problem of reconstructing an orthonormal triad is a subtle task, and if not properly interpreted, one could conceive  a combination of distortion fields and an orthonormal triad that will satisfy Eqs. \eqref{eq::CompCondUsingABCR1212}--\eqref{eq::CompCondUsingABCR2323} but will fail to comply with Eqs. \eqref{eq::deliNj}, \eqref{eq::deliPj}, and \eqref{eq::deliQj}. The failure in the counterexample below results from improperly prescribing compatible information.}

\begin{example}\label{exe::IncompatiblePlanePhase}
{Consider a phase with translation symmetry in the $z$-direction, i.e., $\hatq=\hat{z}$, which requires $b_q=0$,  $\alpha=0$, $\beta=0$, $\gamma=0$, $t=-2\Delta_2$, and $s=2\Delta_1$. Impose, in addition, that $b_p=0$, $t=0$, and $s=s(x)=s_0-s_1x$, where $s_0,s_1\not=0$. The only non-trivial compatibility equations is $\hatn\cdot\nabla s+s^2=0$, which is solved provided that $\hatn=\cos\theta\,\hat{x}+\sin\theta\,\hat{y}$ with $\theta=\theta(x)=\cos^{-1}(s^2/s_1)$. Note, however, that $\nabla\cdot\hatn=\partial_x(\cos\theta)=-2s$. How can deformation modes satisfy the compatibility equations, yet the corresponding reconstructed director does not have the prescribed deformations? The failure is because we did not prescribe, together with the deformation modes, a proper set of 1-forms $\{\eta^i\}$. Indeed, if the reconstructed director had the prescribed modes, then the two sets of 1-forms associated with the frame $\hatn,$ $\hatp=(-\sin\theta,\cos\theta,0)$, and $\hatq=(0,0,1)$ would be $\eta^1=\cos\theta\,\rmd x+\sin\theta\,\rmd y$, $\eta^2=-\sin\theta\,\rmd x+\cos\theta\,\rmd y$, $\eta^3=\rmd z$ and $\eta_1^2=s\eta^2$, $\eta_1^3=s\eta^3$, $\eta_2^3=0$. These differential forms, though, do not satisfy the structure equations. For example, $\rmd\eta^1=\eta^k\wedge\eta_k^1$ does not hold.  On the one hand, $\eta^2\wedge\eta_2^1+\eta^3\wedge\eta_3^1=0$. On the other hand, $\rmd\eta^1=\rmd(\cos\theta\,\rmd x+\sin\theta\,\rmd y)=-\partial_y(\cos\theta)\rmd x\wedge\rmd y+\partial_x(\sin\theta)\rmd x\wedge\rmd y=\partial_x(\sin\theta)\rmd x\wedge\rmd y\not=0$.}
\end{example}

\end{appendices}

\bmhead{Acknowledgments}
This work was funded by the Israel Science Foundation grant no. $1444/21$. L.C.B.dS. acknowledges the support provided by the Mor\'a Miriam Rozen Gerber fellowship for Brazilian postdocs and the Faculty of Physics Postdoctoral Excellence Fellowship.

\section*{Declarations}
The authors declare that they have no conflict of interest.



\begin{thebibliography}{9}

\bibitem{Aminov} Aminov, Yu.: The Geometry of Vector Fields. Gordon and Breach Science Publishers, Amsterdam (2000).

\bibitem{Apostol} Apostol, T.: Calculus, volume II, 2nd Edition, John Wiley and Sons (1969).

\bibitem{B+14} Beller, D. A., Machon, T., \v{C}opar, S., Sussman, D. M., Alexander, G. P., Kamien, R. K., Mosna, R.: Geometry of the cholesteric phase. Phys. Rev. X \textbf{4}, 031050 (2014).

\bibitem{B75} Bishop, R. L.: There is more than one way to frame a curve. Am. Math. Mon. \textbf{82}, 246--251 (1975).

\bibitem{Bryant+91} Bryant, R. L., Chern, S. S., Gardner, R. B., Goldschmidt, H. L., Griffiths, P. A.: Exterior Differential Systems. Springer, New York (1991).

\bibitem{CK19} Chaturvedi, N., Kamien, R. D.: Mechanisms to splay-bend nematic phases. Phys. Rev. E \textbf{100}, 022704 (2019).

\bibitem{C+14} Chen, D., Nakata, M., Shao, R., Tuchband, M. R., Shuai, M., Baumeister, U.,  Weissflog, W., Walba, D. M., Glaser, M. A., Maclennan, J. E., Clark, N. A.: Twist-bend heliconical chiral nematic liquid crystal phase of an achiral  rigid bent-core mesogen. {Phys. Rev. E} \textbf{89}, 022506 (2014).

\bibitem{CK20}  Clelland, J. N., Klotz, T.: Beltrami fields with nonconstant proportionality factor. Arch. Rational Mech. Anal. \textbf{236}, 767--800 (2020). 

\bibitem{dSE21} da Silva, L. C. B., Efrati, E.:  Moving frames and compatibility conditions for three-dimensional director fields. New J. Phys. \textbf{23}, 063016 (2021).

\bibitem{DoCarmo76} do Carmo, M. P.: Differential Geometry of Curves and Surfaces, Prentice-Hall, New Jersey (1976).

\bibitem{DoCarmo94} do Carmo, M. P.: Differential Forms and Applications, Springer Berlin, Heidelberg (1994).

\bibitem{GP95} de~Gennes, P. G., Prost, J.: {The Physics of Liquid Crystals}, 2nd Edition. Oxford University Press, Oxford (1995).

\bibitem{GAE19} Griniasty, I., Aharoni, H., Efrati, E.: Curved geometries from planar director
  fields: Solving the two-dimensional inverse problem. {Phys. Rev. Lett.} \textbf{123}, 127801 (2019). 
  
\bibitem{Kamke1948}
Kamke, E.: Differentialgleichungen L\"{o}sungsmethoden und L\"{o}sungen, Band 1. Gew\"{o}hnliche Differentialgleichungen. 3. Auflage. Chelsea Publishing Company (1948).

\bibitem{LeviCivita} Levi-Civita, T.: The absolute differential calculus, Courier Corporation (1977). 

\bibitem{MA16} Machon, T., Alexander, G. P.: Umbilic lines in orientational order.
Phys. Rev. X \textbf{6}, 011033 (2016).

\bibitem{ME21} Meiri, S., Efrati, E.: Cumulative geometric frustration in physical assemblies. Phys. Rev. E \textbf{104}, 054601 (2021).

\bibitem{Meyer1976} Meyer, R. B.: Structural problems in liquid crystal physics.  In: Les Houches Summer School in Theoretical Physics, vol. XXV, pp. 273--373. Gordon \& Breach, New York (1976).
  
\bibitem{NE18} Niv, I., Efrati, E.: Geometric frustration and compatibility conditions for two-dimensional director fields. Soft Matter \textbf{14}, 424 (2018); Correction: Soft Matter \textbf{14}, 1068 (2018).

\bibitem{PA21} Pollard, J., Alexander, G. P.:  Intrinsic geometry and director reconstruction for three-dimensional liquid crystals. New J. Phys. \textbf{23}, 063006 (2021).  

\bibitem{S18}  Selinger,  J. V.: Interpretation of saddle-splay and the {O}seen-{F}rank free  energy in liquid crystals. Liq. Cryst. Rev. \textbf{6}, 129 (2018). 
 
\bibitem{selingerAnnRev21} Selinger,  J. V.: Director deformations, geometric frustration, and modulated phases in liquid crystals. Annu. Rev. Condens. Matter Phys. \textbf{13}, 49--71 (2022).

\bibitem{selingerPRE22} Selinger, J. V.: Modulated phases of nematic liquid crystals induced by tetrahedral order. Phys. Rev. E \textbf{105}, 024708 (2022).

\bibitem{struik} Struik, D.: Lectures on Classical Differential Geometry. Dover, New York (1961). 

\bibitem{T71} Tenenblat, K.: On isometric immersions of Riemannian manifolds. Bull. Braz. 
Math. Soc. \textbf{2}, 23--36 (1971). 

\bibitem{V19} Virga, E. G.:  Uniform distortions and generalized elasticity of liquid crystals. Phys. Rev. E \textbf{100}, 052701 (2019).

\bibitem{NEcomment} Note that in the original derivation of the $2d$ compatibility equation, \cite{NE18}, the bend was defined as the geodesic curvature of the director integral lines, i.e. with an opposite sign to the more common definition of the bend, which we employ here. Consequently, the equations also differ by a minus sign.

\end{thebibliography}


\end{document}